\documentclass{aastex631}
\usepackage{xcolor}
\usepackage{textgreek}
\usepackage[utf8]{inputenc}
\usepackage[english]{babel}
\usepackage{hyperref}
\hypersetup{
    unicode, 
    colorlinks=true,
    linkcolor=linkcolor,
    citecolor=linkcolor,
    filecolor=linkcolor,
    urlcolor=linkcolor,
}
\usepackage{color,colortbl}
\definecolor{linkcolor}{rgb}{0.0,0.3,0.5}
\usepackage{tensind}
\tensordelimiter{?}
\DeclareGraphicsExtensions{.bmp,.png,.jpg,.pdf}
\usepackage{verbatim}
\usepackage[normalem]{ulem}
\usepackage{orcidlink}
\usepackage{soul}

\usepackage{amsmath,amstext}
\usepackage[T1]{fontenc}
\usepackage{ae,aecompl}
\usepackage[utf8]{inputenc}
\usepackage{enumitem}
\usepackage{bm}
\usepackage{xcolor}

\newcommand{\dhst}{{\rm d_{massive}}}
\newcommand{\tht}{\Theta_1}

\begin{document}

\title{Dwarf Galaxies in the TNG50 Field: connecting their Star-formation Rates with their Environments}

\author{Joy~Bhattacharyya\orcidlink{0000-0001-6442-5786}}
\email{bhattacharyya.37@osu.edu}
\affiliation{The Ohio State University, Department of Astronomy, Columbus, OH 43210, USA}
\affiliation{Center for Cosmology and Astroparticle Physics, 191 West Woodruff Avenue, Columbus, OH 43210, USA}

\author{Annika~H.~G.~Peter\orcidlink{0000-0002-8040-6785}}
\affiliation{The Ohio State University, Department of Physics, Columbus, OH 43210, USA}
\affiliation{The Ohio State University, Department of Astronomy, Columbus, OH 43210, USA}
\affiliation{Center for Cosmology and Astroparticle Physics, 191 West Woodruff Avenue, Columbus, OH 43210, USA}

\author{Alexie Leauthaud\orcidlink{0000-0002-3677-3617}}
\affiliation{Department of Astronomy and Astrophysics, University of California, Santa Cruz, 1156 High Street, Santa Cruz, CA 95064 USA}

\begin{abstract}

The dwarf galaxies comparable to the LMC and SMC, with stellar masses $7.5 <{\rm log}(\mathcal{M}_{\ast}/M_{\odot})<9.5$, are found in a diversity of environments and have long quenching timescales. We need to understand how this phenomenon is connected to the dwarfs' halo properties and their locations in the large-scale structure of the universe. We study the star-formation rates of dwarfs in the TNG50 simulation of the IllustrisTNG project across different environments, focusing on field dwarfs in host halos with virial masses of $9 < {\rm log}(\mathcal{M}_{200}/M_{\odot}) < 11.5$, in contrast to dwarf satellites in hosts with ${\rm log}(\mathcal{M}_{200}/M_{\odot}) \geq 11.5$. Our field dwarf sample is heterogeneous, consisting of primary (central) galaxies, with smaller numbers of secondaries and dwarf galaxies that are on backsplash orbits around massive galaxies. We study how the quenched fraction and star-formation histories depend on the dwarfs' large-scale environment and find that only $\sim 1\%$ of the most isolated dwarfs are quenched. The vast majority of the quenched field dwarfs are backsplash dwarfs located in the neighborhood of cluster-scale halos. We discover a two-halo galactic conformity signal that arises from the tendency of the quenched dwarfs, particularly the backsplash sample, to have a quenched massive galaxy as a neighbor. We attribute the low quenched fractions of the simulated LMC/SMC analogs in the field to the locations of their low-mass hosts in the sparse large-scale environment, which predominate over the relatively small number of backsplash and quenched primary dwarfs in denser environments.

%We quantify the field dwarfs' large-scale environment using the tidal index $\tht$ and the distance to the nearest massive galaxy $\dhst$, and we study how the quenched fraction and star-formation histories are correlated with these metrics. Those with $\dhst>1.5$ Mpc and $\tht<0$ are well isolated, with only $\sim 1\%$ being quenched.

\end{abstract}

%\keywords{Classical Novae (251) --- Ultraviolet astronomy(1736) --- History of astronomy(1868) --- Interdisciplinary astronomy(804)}

\section{Introduction}

The large-scale distribution of matter in the universe is built up in a hierarchical fashion \citep{1978MNRAS.183..341W} according to the standard cosmological paradigm of $\Lambda$ Cold Dark Matter ($\Lambda$CDM). The fluctuations in the primordial density field collapse under gravitational instability, giving rise to halos whose mass is distributed according to a power-law mass function. The halos enable gas to collapse and form stars comprising the galaxies.  Therefore, the properties of galaxies are strongly correlated with the mass of the halos \citep{2018ARA&A..56..435W}. The way that the halos cluster as the universe evolves \citep[e.g.][]{2005MNRAS.363L..66G,2006MNRAS.366....2W,2007MNRAS.374.1303C,2013MNRAS.430.1447K} and how star formation, stellar evolution, stellar and active galactic nuclei (AGN) feedback operate \citep[e.g.][]{1986ApJ...303...39D,2012MNRAS.421.3522H} imply that the properties of galaxy (e.g., color, luminosity, star-formation rate)  scale differently across the range of halo masses. Since the halos are embedded in the large-scale structure of the universe \citep{2012MNRAS.419.2670M,2012MNRAS.419.2133H}, the galaxy properties are influenced by this as well \citep[e.g.][]{2010ApJ...721..193P,2013MNRAS.428.3306W}.

There is a growing interest to understand the connection between a galaxy and its surrounding halo and environment, for dwarf galaxies with masses less than the Milky Way (MW). The halo mass distribution implies that there are abundantly more low-mass halos compared to high-mass ones. Among these, the ones hosting galaxies comparable to the LMC and SMC with stellar masses
$7.5< {\rm log}(\mathcal{M}_{\ast}/M_{\odot})<9.5$ \citep{2002AJ....124.2639V,2006AJ....131.2514H} are not only numerous across the universe but are easily detectable compared to classical and ultra-faint dwarfs (UFDs) \citep{2019ARA&A..57..375S} of even lesser mass. These dwarfs are found in a diversity of environments but are predominantly star-forming \citep{2014MNRAS.442.1396W,2015ApJ...804..136W,2015ApJ...808L..27W,2015MNRAS.447..698P,2024arXiv240414499G}. We are yet to understand if the conditions that disfavor their quenching arise from the baryonic processes internal to their halos or their spatial distribution in the large-scale structure of the universe.

The main focus of relevant studies has been on the satellite dwarfs that are hosted by massive galaxies, particularly MW and its analogs \citep[e.g.][]{1998ARA&A..36..435M,2012AJ....144....4M,2017ApJ...847....4G,2022ApJ...933...47C}. As a result, we now understand that dwarfs with ${\rm log}(\mathcal{M}_{\ast}/M_{\odot})\sim 9$ are quenched through gas depletion, and this occurs over long timescales relative to galaxies with ${\rm log}(\mathcal{M}_{\ast}/M_{\odot})\lesssim 8$, where environmental processes act to quench over shorter timescales \citep{2015ApJ...808L..27W,2015MNRAS.454.2039F,2021ApJ...909..139A,2023ApJ...949...94G}. At the same time, there is an abundance of field dwarfs that exist in isolation from massive galaxies \citep[e.g.][]{2015MNRAS.454.1798K,2018MNRAS.480.3376B}, and are predominantly gas rich and show active star-formation \citep{2012ApJ...757...85G,2015ApJ...809..146B}. Essentially none of the dwarfs with masses $7.5< {\rm log}(\mathcal{M}_{\ast}/M_{\odot})<9.5$ that were situated at distances larger than 1.5 Mpc from a massive galaxy were found to be quenched in the NASA-Sloan Atlas \citep{2012ApJ...757...85G}.

This leads us to the question about what defines the field, and how do we connect the star formation in dwarf galaxies to their halos and large-scale environments? The virial radii of the halos hosting massive galaxies are of the order $\sim 200$ kpc for MW analogs to $\sim 2$ Mpc for the largest cluster. We understand that the gravitational influence of a halo as traced by the splashback radius, extends far beyond its virial radius \citep{2013MNRAS.430.3017B,2014ApJ...789....1D,2014JCAP...11..019A,2015ApJ...810...36M}. This influences the star formation in dwarf galaxies through the processes of environmental quenching. Furthermore, there are hints that the assembly of large-scale structure modulates star-formation across inter-halo scales, with a two-halo conformity effect being observed in massive galaxies \citep{2013MNRAS.430.1447K,2015MNRAS.452.1958H,2023ApJ...943...30O}. This manifests as a correlation between the star-formation rates of the galaxies belonging to halos separated by distance scales of a few Mpc. It is an open question as to which physical processes drive the correlation between star formation in dwarfs and their environments on these diverse set of scales, from the densest to sparsest.

Quenched dwarfs have been found at distances of $\sim 2-4$ virial radii outside massive galaxies \citep{2009ApJ...692..931L,2009ApJ...697..247W,2014MNRAS.444.2938H}, and we believe that many of these are `backsplash' galaxies. These galaxies experienced pericentric passages within the virial radii of massive galaxies, which can quench them \citep{2005MNRAS.356.1327G,2014MNRAS.439.2687W}, and they subsequently orbit out into the field. Backsplash dwarfs have been seen in the Local Group \citep{2012MNRAS.426.1808T} and beyond \citep{2023arXiv231200773B}. There exist other mechanisms of environmental quenching that process infalling dwarfs before they cross the virial radius of a massive galaxy, in what is collectively known as pre-processing \citep{2004PASJ...56...29F,2017MNRAS.467.3268R}. 

The groups of dwarf galaxies serve as key sites for pre-processing prior to their accretion onto a massive galaxy \citep{2023MNRAS.525.3849S}. For example, the LMC/SMC and its satellites belonged to a dwarf group that merged with the satellite population of MW after its accretion \citep{2008ApJ...686L..61D,2015MNRAS.453.3568D,2022ApJ...940..136P}. These groups are expected given the hierarchical distribution of matter \citep{2013MNRAS.428..573S} and we continue to discover many of these in different environments. The dwarf groups that exist in the field have been referred to as `associations' \citep{2006AJ....132..729T,2018MNRAS.480.3376B}, e.g., the NGC 3109 group \citep{2013A&A...559L..11B,2015ApJ...812L..13S}. These systems are ideal to study how quenching may proceed in low-mass groups \citep{2019ApJ...886..109C,2020MNRAS.492.1713G,2022arXiv220909262G,2021MNRAS.500.3854D,2024arXiv240903999D,2022MNRAS.513.2673J,2022MNRAS.514.5276S}.

%Therefore, in the low density environment of the field, most of the halos are found to be low mass halos \citep{2012MNRAS.419.2133H} and are assembled at later redshifts \citep{2024arXiv240113252F}.

There has been an emphasis to survey LMC/SMC analogs across different environments and study their host halos, including the satellite populations that they contain \citep[e.g.][]{2015ApJ...805....2S,2016ApJ...828L...5C,2017ApJ...847....4G,2022ApJ...933...47C,2024ApJ...974..273M}. For example, the Merian Survey \citep{2023arXiv230519310L,2024arXiv241001884D} of star-forming dwarfs in the mass range $8< {\rm log}(\mathcal{M}_{\ast}/M_{\odot})<9$ will provide a sample size of $\sim 85,000$ that will allow studies of their dark matter halos using weak lensing \citep{2020PDU....3000719L}. Weak lensing from photometric data of dwarfs in the Dark Energy Survey \citep{2023arXiv231114659T} has provided constraints on their halo mass. These measurements of the dark matter, along with star formation activity in the dwarfs, will further constrain the galaxy-halo connection. We need to determine how the star-formation histories (SFH) of the dwarfs give rise to this connection by using hydrodynamical simulations.

To answer these questions, we undertake a study of massive dwarfs as a function of environment with the TNG50 simulation of the IllustrisTNG project \citep{2019ComAC...6....2N,2019MNRAS.490.3196P}, which is ideal given its fine mass resolution coupled with large box size. We closely build on the analysis of \citet{2021MNRAS.508.1652J} and \citet{2024ApJ...961..236C} to place in context the star-formation histories of simulated field dwarf galaxies with their large-scale environments, with additional emphasis on the backsplash dwarfs and the dwarf associations. We focus on the field dwarf galaxies residing in low-mass host halos. We identify the quenched and star-forming sub-populations among them, and study how the extremities of star-formation rates depend on estimators that trace their large-scale environmental density. The details of the simulation are found in Section \ref{sec:simulation}. We then characterize dwarf galaxies' star-formation rates as well as their large-scale environments in Section \ref{sec:definitions}. This lets us identify a population of field dwarf galaxies, including  backsplash dwarfs. We then proceed to study the baryonic content, quenched fractions, stellar assembly history and galactic conformity of this sample in Section \ref{sec:results}. We place our findings in context with the contemporary understanding of dwarfs and propose future observations in Section  \ref{sec:discussion}.

\section{Simulation} \label{sec:simulation}

We use the cosmological magneto-hydrodynamical simulation IllustrisTNG project for our analysis \citep{2018MNRAS.475..624N,2018MNRAS.475..676S,2019MNRAS.490.3196P}. These are run using the moving-mesh code ${\rm AREPO}$ \citep{2010MNRAS.401..791S} and contain realistic prescriptions for baryonic physics that include star formation, stellar evolution, gas heating, gas cooling, growth of black holes, and AGN feedback \citep{2017MNRAS.465.3291W,2018MNRAS.473.4077P}. These simulations assume a flat $\Lambda$CDM cosmology with parameters adopted from the \citet{2016A&A...594A..13P} results: $h = 0.6774, \Omega_m = 0.3089, \Omega_{\lambda} = 0.6911, \Omega_b = 0.0486, n_s = 0.9677, \sigma_8 = 0.8159$. There are three boxes, TNG300, TNG100 and TNG50, with progressively increasing mass resolution and decreasing volume. There are 100 snapshots for each run, approximately equally spaced in redshift between $z\sim20$ to $z=0$.

Among the three boxes, we deem the TNG50 simulation \citep{2019ComAC...6....2N,2019MNRAS.490.3196P}, spanning $35~{\rm Mpc\ h^{-1}}$ along each side, as ideal for studying dwarf galaxies in a $z=0$ universe. This is because it has a baryonic mass resolution of $m_{\rm baryon} \sim 8.5 \times 10^4 M_{\odot}$ and a dark matter resolution of $m_{\rm DM} \sim 4.5 \times 10^5 M_{\odot}$, that is necessary to resolve galaxies down to $\gtrsim 100$ star particles. There are $2160^3$ gas cells and provides an average spatial resolution of star-forming ISM gas of $\sim 100-200$ pc. We choose TNG50 over the TNG100 and TNG300 because we prioritize resolving down to dwarfs with stellar mass ${\rm log}(\mathcal{M}_{\ast}/M_{\odot})=7.5$ over a large simulation volume. 

The simulated galaxies in TNG50 live in subhalos that are identified using the ${\rm SUBFIND}$ algorithm \citep{2001MNRAS.328..726S}. This algorithm is based on the friend-of-friend (FoF) prescription that uses a linking length of $b = 0.2$ times the mean inter-particle distance. It assigns subhalos in a hierarchical manner with all the subhalos assigned to a FoF group belonging to the parent dark matter halo, which we shall refer to as the host halo henceforth. We define the virial radii of these hosts to be $\mathcal{R}_{200}$ which is the comoving radius of the sphere within which the mean density is 200 times the critical density of the universe. Similarly, the mass $\mathcal{M}_{200}$, defined as the total mass contained within $\mathcal{R}_{200}$, is adopted as the host halo mass corresponding to a FoF group. The virial radius $\mathcal{R}_{200}$ however does not capture the size of the full galaxy distribution within the FoF group.  For this analysis, we consider a ``satellite" or ``secondary" galaxy to be any non-central galaxy assigned to the FoF group, regardless of whether it lies within the virial radius or not.  However, we consider an alternative definition of field galaxies with respect to the massive hosts (see Appendix \ref{sec:outside_halo}), defining a satellite galaxy to be within $\mathcal{R}_{200}$.  We found that our results are robust across definitions.

All of the relevant baryonic quantities, including the stellar mass $\mathcal{M}_{\ast}$, gas mass $\mathcal{M}_{\rm gas}$ and star-formation rate (SFR), are computed from star/gas particles located within twice the radius which contains half of the stellar mass of the subhalo. However, in circumstances when we are dealing with the dynamical mass, we use the total mass of a subhalo  $\mathcal{M}_{\rm dyn}$ \citep{2021MNRAS.500.3957E}. This is the total mass of the subhalo comprising all the dark matter, star, gas, and black hole constituents. We refer to the most massive subhalo in the host halo as the primary, or occasionally the central, and its companion subhalos of lesser mass in the same host halo as secondaries.

\section{Definitions} \label{sec:definitions}

\subsection{Dwarf Galaxy} \label{sec:dwarf_def}

\begin{figure}
\centering
\includegraphics[scale=0.375]{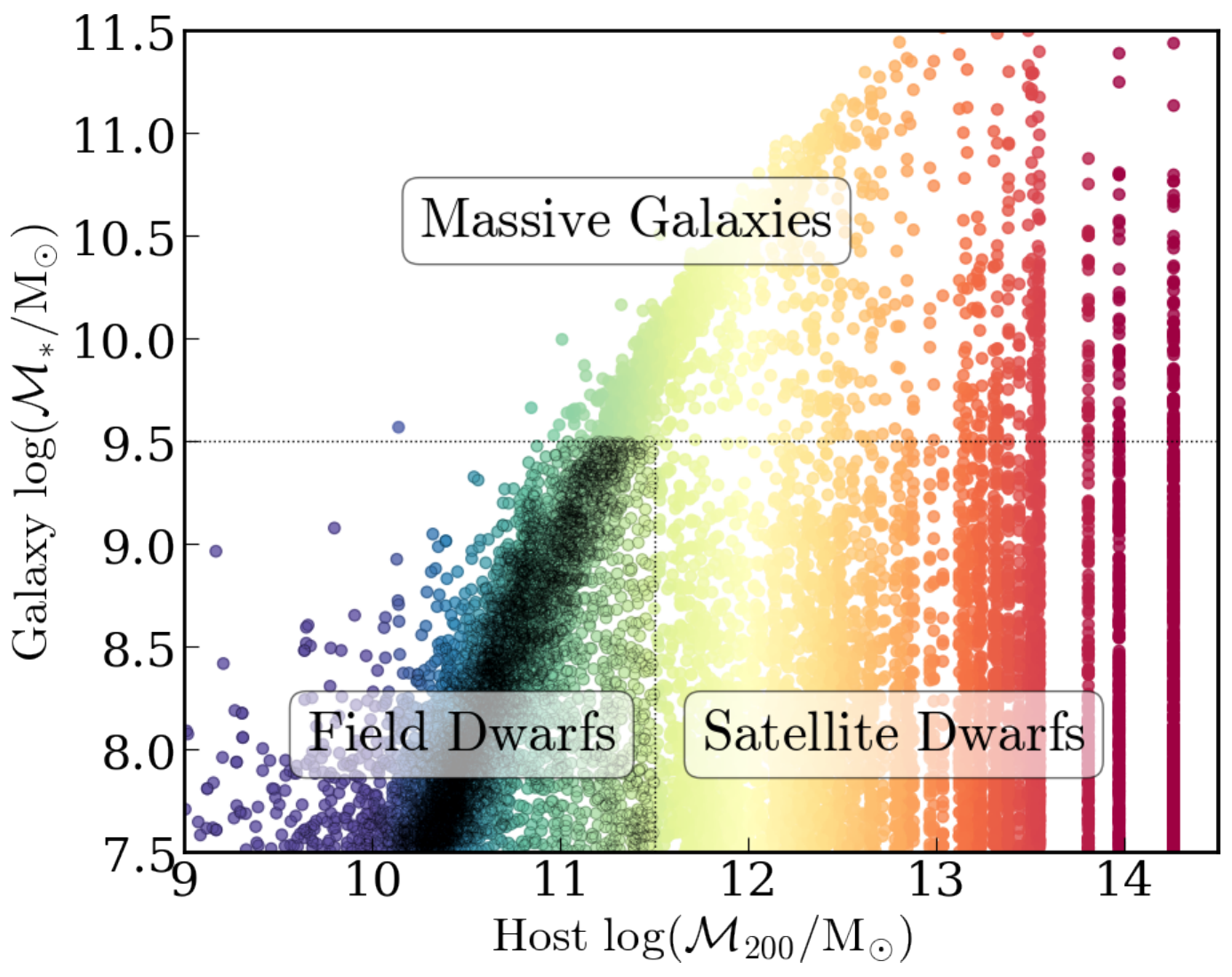}
\caption{The scatter plot of the SUBFIND-assigned stellar mass $\mathcal{M}_{\ast}$ vs. host virial mass $\mathcal{M}_{200}$ for all the galaxies in the box of TNG50. Each point is colored according to its host virial mass. We segment this space according to the samples that we refer to in this work. The upper half represents the massive galaxies with ${\rm log}(\mathcal{M}_{\ast}/M_{\odot})>9.5$ with respect to which we define the density estimators in Sec. \ref{sec:env_sfr} and the backsplash galaxies in Sec. \ref{sec:backsplash}. The lower half represents the dwarfs in the simulation with $7.5< {\rm log}(\mathcal{M}_{\ast}/M_{\odot})<9.5$ and has been divided into two segments. The lower right segment represents the satellite dwarfs of the FoF-assigned host halos with ${\rm log}(\mathcal{M}_{200}/ M_{\odot})\geq 11.5$, while the lower left segment represents our target sample of field dwarfs with ${\rm log}(\mathcal{M}_{200}/ M_{\odot})< 11.5$.  \label{fig:hostM_subhaloM}}
\end{figure}

To define what constitutes a dwarf in the simulation, we choose galaxies in the range of $7.5< {\rm log}(\mathcal{M}_{\ast}/M_{\odot})<9.5$. The lower limit is determined by the baryonic mass resolution of the simulation. It was demonstrated by \citet{2021MNRAS.500.4004D} with the TNG100 and TNG300 runs that poor baryon resolution leads to a higher quenched fraction in the galaxies with $<100$ star particles. For the baryon resolution of TNG50, ${\rm log}(\mathcal{M}_{\ast}/M_{\odot})=7$ corresponds to a galaxy with a little over 100 star particles, with this being the same cut adopted in \citet{2021MNRAS.506.4760D}. However, \citet{2021MNRAS.508.1652J} showed by comparing to lower-resolution runs of the TNG50 simulations (see their Appendix A)  that the cumulative SFHs are poorly converged for $7 < {\rm log}(\mathcal{M}_{\ast}/M_{\odot}) \lesssim 7.5$. These runs of TNG50-2, TNG50-3 and TNG50-4 show that the primary dwarfs at these scales build up their stellar masses earlier on account of the efficiency of supernovae feedback at these resolutions. Since we are particularly interested in star formation in dwarfs, most of which are primaries, we adopt ${\rm log}(\mathcal{M}_{\ast}/M_{\odot})=7.5$ as the lower limit on the stellar mass of a dwarf galaxy in our sample. Given the baryonic mass resolution of TNG50, this corresponds to $\geq 370$ star particles for this sample of dwarf galaxies. The upper cut-off approximately represents the mass of LMC-analogous dwarf since the stellar mass of LMC is estimated to be ${\rm log}(\mathcal{M}_{\ast}/M_{\odot}) = 9.43$ \citep{2002AJ....124.2639V}.

In Fig. \ref{fig:hostM_subhaloM}, we plot the galaxy stellar mass $\mathcal{M}_{\ast}$ against their parent halo virial mass $\mathcal{M}_{200}$. Here we notice the prominent sequence made up of the primaries for each of the hosts, and the galaxies below this sequence are the respective secondaries. Based on the convention in the previous section, those galaxies with ${\rm log}(\mathcal{M}_{\ast}/M_{\odot})\geq 9.5$ are `massive', comparable or larger than the MW, and make up the upper half of this space. The dwarf galaxies with $7.5< {\rm log}(\mathcal{M}_{\ast}/M_{\odot})<9.5$ that are the emphasis of this work make up the lower half of this space.

We find that there is a great diversity in the types of host halos that the simulated dwarfs occupy, based on the distribution in Fig. \ref{fig:hostM_subhaloM}. There are dwarfs that are assigned by the SUBFIND FoF algorithm as satellites of massive host halos, which occupy the lower right half of the plot, and those that are isolated from the latter, in the lower left half. We choose ${\rm log}(\mathcal{M}_{200}/ M_{\odot})= 11.5$ as a boundary for the two types of systems because this approximately represents the edge of the central sequence at fixed ${\rm log}(\mathcal{M}_{\ast}/M_{\odot})=9.5$. On the other hand, the lower limit of ${\rm log}(\mathcal{M}_{200}/ M_{\odot})= 9$ is chosen to account for the dwarfs that have been tidally stripped, and as a result, possess reduced halo masses compared to the typical dwarf of ${\rm log}(\mathcal{M}_{\ast}/M_{\odot})=7.5$ on the central sequence. Therefore, this defines the selection of the `field' sample --- dwarfs that are spatially isolated from massive galaxies in groups and clusters, that are the particular emphasis of this work. In contrast to this, there is the `satellite' dwarf sample situated in the FoF assigned host halos of ${\rm log}(\mathcal{M}_{200}/ M_{\odot})> 11.5$ that we additionally refer in this work.

\subsection{Dwarf Primaries \& Secondaries}

We are able to further subdivide this field dwarf sample based on how they occupy their host halos. Firstly we notice two subsamples:  the primary dwarfs which are the central galaxies in the host halos, and the secondary dwarfs, which are less massive than the primaries. The most numerous among them are the primaries. Our definition of dwarf galaxies ($7.5 <{\rm log}(\mathcal{M}_{\ast}/M_{\odot})<9.5$) in the field ($9 < {\rm log}(\mathcal{M}_{200}/M_{\odot}) < 11.5$) yields 5843 members, of which 5003 are primaries, 465 are secondaries and 375 are backsplash dwarfs. While the backsplash dwarfs by way of our definition are also classified as primaries, we have assigned them their own class due to their unique orbital history.

There are 429 primaries which have secondary subhalos in their same host halo. We collectively refer to the whole class of the secondary dwarfs because most of them are in pairs with a primary i.e., that the vast majority of dwarf groups only have two members, down to the resolution limit of our study. However, we note that some of the field dwarf host halos have more than two subhalos, numbering between 3 to 6. We find that out of the 505 secondaries, 364 (72\%) are the second-most massive, 56 (11\%) are the third-most massive, 7 (1\%) are the fourth-most massive and 2 ($<1$\%) are the fifth-most massive. Our system of classification closely resembles the one in \citet{2021MNRAS.508.1652J}, albeit they classify their sample as centrals, group members and backsplash dwarf galaxies. 

\subsection{Backsplash Dwarfs} \label{sec:backsplash}

\begin{figure*}
\centering
\includegraphics[scale=0.4]{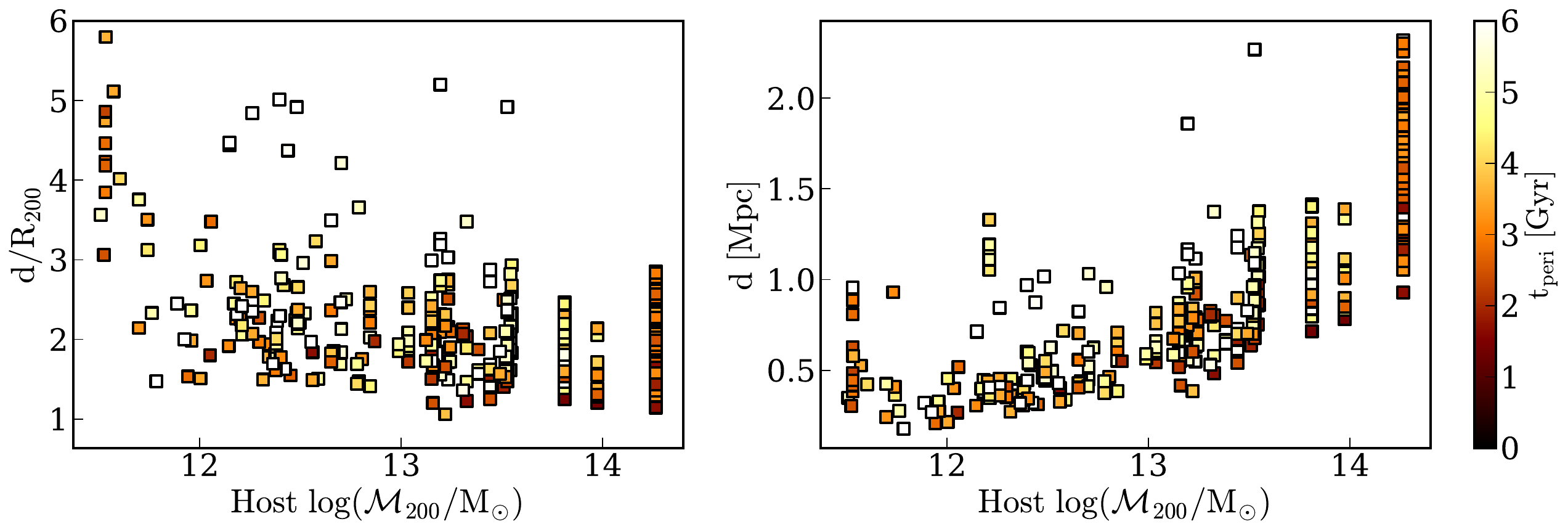}
\caption{ The radial distribution of backsplash dwarfs with present day distances normalized with respect to the virial radius of the massive halo where they had their last pericentric passage (\textit{left}), and present day distances in units of Mpc (\textit{right}). See Fig. 7 in \citet{2021NatAs...5.1255B} for comparison. \label{fig:backspl_rads}}
\end{figure*}

Among this field dwarf sample, 6.4\% are backsplash galaxies that have previously orbited inside a massive host halo. We determine the backsplash dwarfs by closely following the prescription of \citet{2023MNRAS.520..649B}. For each SUBFIND assigned host with ${\rm log}(\mathcal{M}_{200}/ M_{\odot})\geq 11.5$ in the $z=0$ snapshot, we search for the backsplash candidate subhalos that are located outside their virial radius at host-centric distances $R_{200}< r < 10R_{200}$. We use the most massive subhalo inside this host to trace the evolution of the host's center and its virial radius. To reduce the exchange of backsplash dwarfs \citep{2022MNRAS.514.3612N} between the hosts or the interaction with more than one host, we reject hosts whose centers shift by $>8$ Mpc Gyr$^{-1}$ with respect to the fixed box across two snapshots. We query the main progenitor branches of the merger trees of the host's most massive subhalo and the backsplash candidate subhalos. We reject all those branches whose main progenitor branch does not extend beyond 40 snapshots, i.e., $z\geq 0.7$, to ensure that they correspond to the massive and backsplash dwarf galaxies that are stable over time. Finally, if the progenitor of the backsplash candidate makes an appearance within $r < R_{200}$ of the progenitor host at some $z>0$, we select that as a backsplash dwarf galaxy. 

In Fig. \ref{fig:backspl_rads}, we plot along the vertical axes the $z=0$ distances of the backsplash dwarfs from the centers of the halos where they performed their most recent pericentric passages, in units of the latter's virial radii in the left panel, and Mpc in the right panel. The horizontal axes show the $\mathcal{M}_{200}$ of the massive host halo inside which the backsplash dwarf performed a pericentric passage. The points are colored according to the lookback time corresponding to when they did their most recent pericentric passages near the massive host center. Refer to the plots in Fig. 7 of \citet{2021NatAs...5.1255B} for comparison since those show the normalized and physical host-centric distances of ultra-diffuse dwarfs (UDG) in the same TNG50 simulation. We find a median $t_{\rm peri} = 3.6$ Gyr with a minimum and maximum of 1.3 Gyr and 9.5 Gyr, respectively. The backsplash dwarfs are found at distances of $\sim 0.5-2$ Mpc away from the massive host halos, with an increase with respect to $\mathcal{M}_{200}$, which is expected given the fact that the sizes of the halos, e.g., their virial radii $R_{200}$ are proportional to $\mathcal{M}_{200}$. 

In the Appendix \ref{sec:outside_halo}, we test the robustness of our selection of field dwarfs by considering an extended sample that takes into account not only those with host masses $9 < {\rm log}(\mathcal{M}_{200}/M_{\odot}) < 11.5$ but also those with which are situated beyond the virial radius of their massive host halos (${\rm log}(\mathcal{M}_{200}/ M_{\odot})\geq 11.5$), i.e.,host-centric distances $d>R_{200}$. This adds to the existing sample, 882 backsplash and 825 primary dwarfs that are outside the virial volume of the massive hosts yet are assigned to the same Fof group by SUBFIND as the latter. We find that the results are qualitatively unchanged even if some quantitative measurements are. 

%These massive hosts have masses ${\rm log}(\mathcal{M}_{200}/M_{\odot}) \sim 12-13$.

\subsection{Star-formation \& Quenching} \label{sec:sfr_quench}

\begin{figure}
\centering
\includegraphics[scale=0.375]{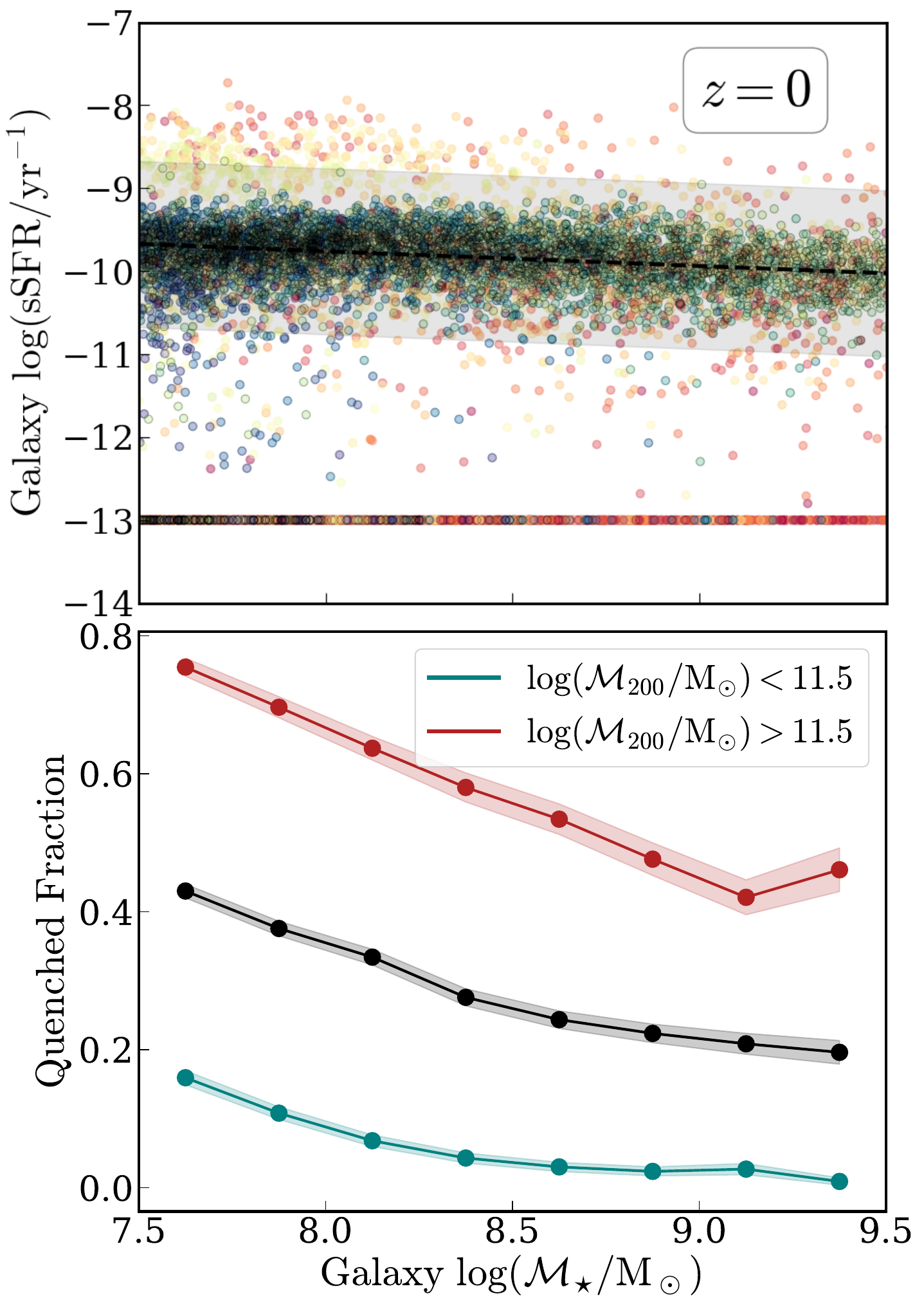}
\caption{ \textit{Top:} Scatter plot of sSFR-$\mathcal{M}_{\ast}$ for all galaxies with $7.5< {\rm log}(\mathcal{M}_{\ast}/M_{\odot})<9.5$. in the simulation volume. The color scheme used in Fig. \ref{fig:hostM_subhaloM} is adopted. The dashed black line represents the SFMS that we derive from the median sSFR in bins of $\mathcal{M}_{\ast}$. The shaded gray region is the $\pm 1$ dex region about the SFMS and any galaxy below this is deemed quenched. The galaxies with sSFR below the limit of ${\rm log\ sSFR/yr^{-1}}=-13$ have been assigned this value and are shown near the bottom of the plot. \textit{Bottom:} The quenched fraction of the galaxies as a function of $\mathcal{M}_{\ast}$ in two bins of host mass. The teal one represents the field dwarfs $9< {\rm log}(\mathcal{M}_{200}/ M_{\odot})< 11.5$ while the red one represents their satellite counterparts in high-mass halos with  ${\rm log}(\mathcal{M}_{200}/ M_{\odot})>11.5$. The field dwarfs are more star-forming than those in high-mass hosts.  \label{fig:sfnms_geha12}}

\end{figure}

%The centrals with $\mathcal{M}_{\ast}>10^{10} M_{\odot}$ have outliers in their SFR with respect to the SFMS, due to internal quenching (CITE) as a result of which these are left out.

We seek to study how the star-formation rates of the dwarf galaxies depend on their respective environments. For this purpose, we first need to identify the star-forming main sequence (SFMS), then study deviations therefrom. We look at the full diversity of star-formation rates about the SFMS with an emphasis on quenching, i.e., star-formation rates below the SFMS, and study the dependence of this as a function of the dwarf's stellar mass and environment. In the upper panel of Fig. \ref{fig:sfnms_geha12}, we plot the specific star-formation rate (sSFR=SFR/$\mathcal{M}_{\ast}$) as a function of $\mathcal{M}_{\ast}$ for all  dwarfs in the range $7.5< {\rm log}(\mathcal{M}_{\ast}/M_{\odot})<9.5$. The points are colored according to the color scheme in Fig. \ref{fig:hostM_subhaloM}. In order to determine the SFMS, we calculate the median sSFR for dwarfs in 0.2 dex bins of ${\rm log}(\mathcal{M}_{\ast}/M_{\odot})$, following which we use linear regression on the median estimates. The SFMS extrapolated across the full range of stellar masses is shown using the dashed black line in the same figure. The gray shaded region represents the $\pm 1$ dex region around the SFMS. 

To determine whether a galaxy has been quenched, we use the definition used in \citet{2021MNRAS.506.4760D}, wherein dwarf galaxies which lie below 1 dex from the SFMS are deemed quenched. From a preliminary inspection, we find that most of the field dwarf galaxies, shown using the green points lie within $\pm 1$ dex of the SFMS, showing that this sample is largely star-forming. To explore this further, we plot the quenched fraction of the dwarfs in bins of stellar masses, as seen in the lower panel of Fig. \ref{fig:sfnms_geha12}. The details and implications are discussed in Sec. \ref{sec:quench_frac_mstar}, after we introduce how we quantify the large-scale environment of dwarfs.

\subsection{Environment} \label{sec:env_sfr}

Having characterized the dwarf SFMS, we focus on the subset of field dwarf galaxies that we have selected using their $\mathcal{M}_{\ast}$ and $\mathcal{M}_{200}$. To identify the `field' in the simulated volume, we need to carefully establish what constitutes the environment and accordingly choose a tracer that best estimates this. According to \citet{2012MNRAS.419.2670M}, there are two types of environments --- the \textit{large-scale environment} that identifies where the galaxies are situated with respect to the large-scale structure of the universe, and the \textit{local environment}, which corresponds to the substructure of the parent halo itself. Since in this work we are mainly interested in the field dwarf galaxies, we need to use density estimators that best identify the galaxy density at the scales of intra-halo separation i.e., the large-scale environment.

In order to identify the neighboring galaxies around a dwarf we use \texttt{scipy.kdtree} to construct a 3-dimensional tree and query galaxies from it. We define a fiducial density estimator, the overdensity parameter ${\rm log}(1+\delta_5)$, frequently employed in literature to define both the large-scale and local environments of galaxies \citep[e.g.][]{2012MNRAS.419.2670M,2013MNRAS.428.3306W}. We then define the proximity to a massive galaxy $\dhst$ and the tidal index $\tht$ as density estimators that we apply to study the large-scale environments for our sample of dwarfs. These two estimators represent observables that are measurable from galaxy surveys, which we discuss further in the sub-sections.

%While the use of these density estimators is central to our study of field dwarf galaxies, we do not explicitly place cuts of them in order to derive the sample of field dwarfs. While the formalism of using a isolation criteria is commonly used, we present an alternative formalism. This is because the host halo mass also traces the large-scale environment, with smaller independent halos being preferably found in fields \citep{2012MNRAS.419.2133H}. Since we select the field dwarfs according to their host halo mass which correlates with the large-scale environment already, this renders the adoption of any cut on ${\rm log}(1+\delta_5)$, $\dhst$ and  $\tht$ as trivial.

\subsubsection{Halo Mass \& Overdensity Parameter} \label{sec:halomass_delta5}

\begin{figure*}
\centering
\includegraphics[scale=0.375]{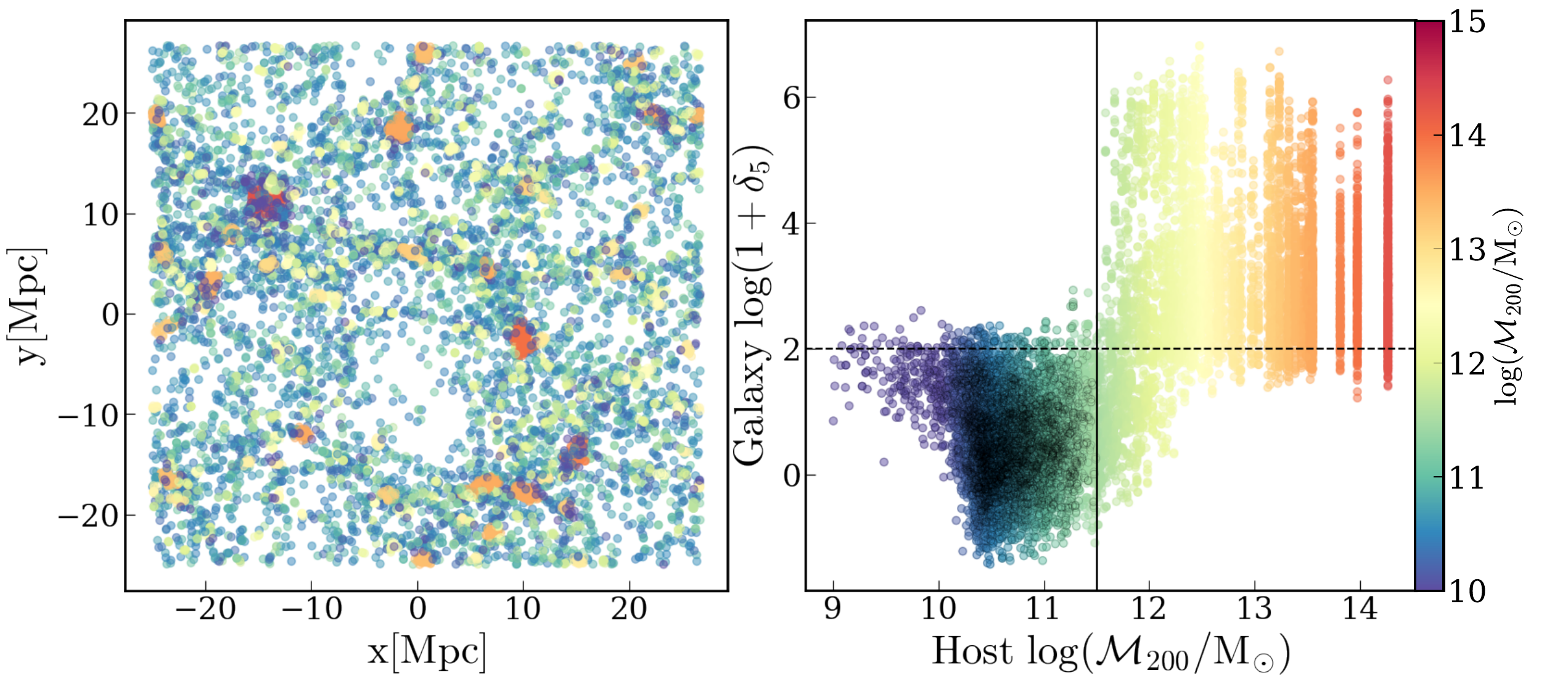}
\includegraphics[scale=0.375]{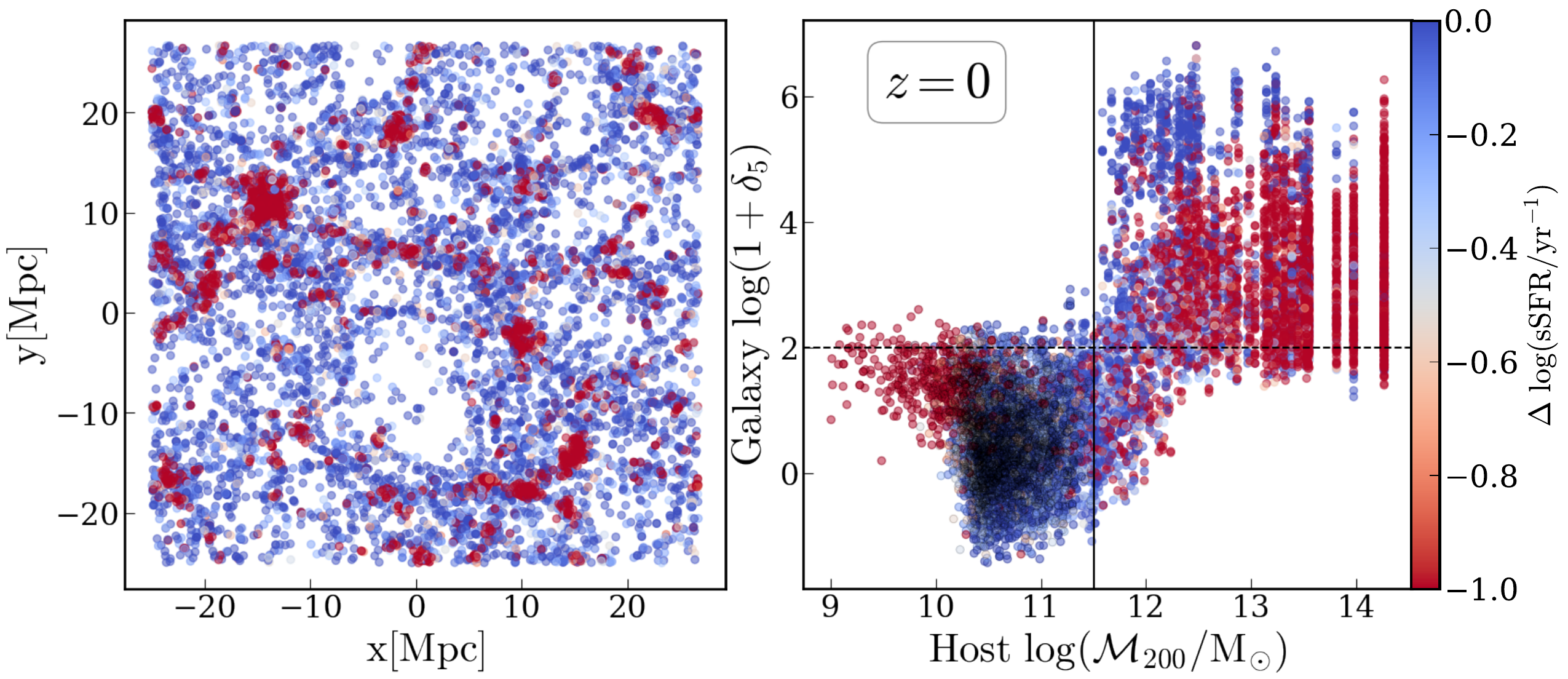}
\caption{ \textit{Upper left:} The TNG50 simulation projected along the z-axis of the simulated box, showing all the galaxies with ${\rm log}(\mathcal{M}_{\ast}/M_{\odot})\geq 7.5$, with each point colored according to the host halo $\mathcal{M}_{200}$ they inhabit. The galaxies in cluster or group scale halos shown in red-orange-yellow colors tend to inhabit nodes, while those corresponding to the field dwarfs shown in blue-green are in filaments and voids. \textit{Upper right:} The distribution of all of the galaxies in  $\mathcal{M}_{200}-{\rm log}(1+\delta_5)$ space, where the vertical solid line defines the limit for the host's $\mathcal{M}_{200}$ used to define the field dwarf sample in Sec. \ref{sec:definitions}. The relevant sample occupies the lower left corner of this plot, showing that with these cuts we are effectively able to select a field dwarf sample with ${\rm log}(1+\delta_5)\lesssim 2$. \textit{Lower:} Same as panel above except that the points are colored according to the galaxy's sSFR deviation from the SFMS, with bluer colors indicating active star-formation as opposed to saturated red color implying a quenched state. \label{fig:dens_map}}
\end{figure*}

Here we demonstrate how our sample of dwarfs, selected according to their host mass $\mathcal{M}_{200}$, is essentially a field sample. In the upper left panel of Fig. \ref{fig:dens_map}, we show the X-Y projection of the simulation volume of TNG50. Each point represents a galaxy with ${\rm log}(\mathcal{M}_{\ast}/M_{\odot})\geq 7.5$, with their colors representing their host halo mass $\mathcal{M}_{200}$. The red/orange points are the satellite dwarfs that reside in the cluster or group environments with ${\rm log}(\mathcal{M}_{200}/M_{\odot})>11.5$. These are the dense nodes of the cosmic web, as seen in Fig. \ref{fig:dens_map}. The blue/green points are the field dwarfs in halos with $9<{\rm log}(\mathcal{M}_{200}/M_{\odot})<11.5$ that comprise the sample of interest in this work. These are distributed throughout the volume, mainly the voids and filaments. We place this observation in context with the results of
\citet{2012MNRAS.419.2133H,2013MNRAS.428.3306W}, who showed how low-mass halos lived in low-density environments.

As a fiducial measure of the large-scale environment of our dwarf sample, we use the overdensity parameter \citep{2005ApJ...634..833C,2010ApJ...708..505K,2010ApJ...721..193P,2013MNRAS.428.3306W} that is defined as:

\begin{equation}
(1+\delta_5) = \frac{18}{4\pi \bar{n}} \frac{1}{(d_{5NN})^3}.
\end{equation}

This is the density of the 6 galaxies contained within the distance to the 5th nearest neighbor $D_{5NN}$ of a galaxy, contrasted to the mean number density of galaxies $\bar{n}=0.1 $ Mpc$^{-3}$. For identifying the nearest neighbors and calculating the mean density, we use all the galaxies with ${\rm log}(\mathcal{M}_{\ast}/M_{\odot})\geq 7.5$ in the entire TNG50 box. The mean density is calculated by counting the total number of galaxies above the threshold and then dividing it by the volume of the box.

In the upper right panel of Fig. \ref{fig:dens_map}, we plot the overdensity parameter as a function of the host halo mass adopted in this work. We find that there are clearly two regimes that arises from the halo occupation distribution rising above unity at ${\rm log}(\mathcal{M}_{200}/M_{\odot}) \sim 12 $. The lower left corner of the space is occupied by the field dwarf sample of our interest, whereas the upper right corner represents the massive galaxies and their dwarf satellites. This justifies our assertion that by selecting the galaxies having hosts with $9<{\rm log}(\mathcal{M}_{200}/M_{\odot})<11.5$, we are essentially selecting them by their large-scale environment corresponding to ${\rm log}(1+\delta_5)\lesssim 2$, which in this case is the field. In the Appendix \ref{sec:outside_halo}, we test the robustness of our key results by considering an extended sample of dwarfs that incorporates the backsplash and infalling dwarfs beyond $R_{200}$ but belonging to the SUBFIND-assigned massive halos. Although this leads to a larger scatter in the density estimators in the densest regions, the general trends between environment and star formation remain unchanged.

In the lower panel of Fig. \ref{fig:dens_map}, we show the same plots as those in the upper panel except that here the points are colored according to the deviation of the galaxy's sSFR from the SFMS in the range of $-1<\Delta({\rm log sSFR})<0$ to visually depict the gradient in sSFR between the SFMS and the quenching threshold. The bluer points are all star-forming dwarfs and located close to the SFMS while the redder points show those dwarfs which are close to being quenched or are quenched already. Any dwarf with an sSFR offset beyond this range will be at the saturated ends. Examining the plots we find that the quenched galaxies are typically located in the densest region of the box and that the field dwarfs inside the low-mass halos are predominantly star-forming except for a few that have masses ${\rm log}(\mathcal{M}_{200}/M_{\odot})\lesssim 10$. We investigate the dependence of star-formation on large-scale enviornment further in Sec. \ref{sec:quench_frac}.

%Nonetheless, there is still a wide range of densities present and we devise estimators that capture them better in the following sections.

\subsubsection{Massive Galaxy Proximity} \label{sec:dhost}

A simple and direct way to trace the large-scale environment of a field dwarf is to measure the distance to the nearest massive or luminous galaxy. For example, the studies undertaken in \citet{2012ApJ...757...85G} and \citet{2009ApJ...697..247W} use physical projected distances and projected distances normalized by the virial radii of the hosts, respectively. \citet{2012ApJ...757...85G} quantifies the environment of a sample of dwarf galaxies with $7<{\rm log}(\mathcal{M}_{\ast}/M_{\odot})<9$ drawn from SDSS DR8 with the distance to the nearest luminous galaxy ${\rm d_{host}}$. The luminous galaxies are drawn from a 2MASS sample, which is complete within $M_{Ks} < -23$. Assuming a mean stellar mass-to-light ratio of unity, this limit corresponds to massive galaxies with stellar mass of $\gtrsim 2.5 \times 10^{10} M_{\odot}$. Furthermore, based on the quenched fraction of dwarfs, \citet{2012ApJ...757...85G} identifies field dwarfs as those devoid of a neighboring massive galaxy within 1.5 Mpc, i.e., dwarfs with $\dhst>1.5$ Mpc. The same estimator was further used to explore the quenching of dwarf galaxies in \citet{2014MNRAS.442.1396W,2024ApJ...961..236C}.

In this work, we measure 3D distances between the dwarf and massive galaxies in TNG50 simulation. For each member galaxy of the low-mass halo sample, we find the nearest neighboring massive galaxy with ${\rm log}(\mathcal{M}_{\ast}/M_{\odot}) \geq 9.5$ and we refer to this density estimator as $\dhst$. While we note that this mass threshold is almost an order lower than the massive galaxy selection in \citet{2012ApJ...757...85G}, we have chosen to stay consistent with the massive galaxy selection outlined in Sec.~\ref{sec:dwarf_def}. The results that we present in this work are in agreement with \citet{2012ApJ...757...85G} and we have tested that our conclusions are robust to the definition here.

%In the lower left panel of Fig. \ref{fig:dens_map} we show the dependence of $\dhst$ on the overdensity parameter $\delta_5$. The horizontal axis is in logarithmic scale and the points are colored by the host halo mass. The more isolated galaxies have a high $\dhst$ and low $\delta_5$ and vice versa. 

\subsubsection{Tidal Index} \label{sec:tidalind}

A frequently employed measure of the environmental density of a galaxy is the dimensionless tidal index $\tht$ \citep{1999IAUS..186..109K,2004AJ....127.2031K,2013AJ....145..101K,2018MNRAS.479.4136K,2018MNRAS.480.3376B,2024ApJ...966..188M}. The most crucial aspect of this estimator is that it takes into account the dynamical masses of neighboring galaxies. The Main Disturber (MD) is defined as the galaxy that exerts the maximum tidal force among the neighboring galaxies and $\tht$ is proportional to this force. The tidal index for the $j^{th}$ galaxy is calculated as:

\begin{equation}
\Theta_{1,j} = {\rm max \biggl\{ log}\biggl(\frac{\mathcal{M}_{{\rm dyn},k}/M_{\odot}}{(d_{jk}/{\rm Mpc})^3}\biggl) \biggr\}_{k=1,..,5} - C
\end{equation}

The constant $C = 10.96$ is chosen in \citet{2013AJ....145..101K} based on dynamical arguments such that galaxies with $\tht=0$ are situated on the zero-velocity sphere of its MD.  A galaxy with a low and negative $\tht$ is located in a low-density environment and vice versa. In that case, $\tht<0$ would imply that the galaxy is isolated, whereas a galaxy with $\tht>0$ is gravitationally coupled to the MD galaxy. We determine the MD from the 5 nearest neighboring massive galaxies with ${\rm log}(\mathcal{M}_{\ast}/M_{\odot})\geq 9.5$ as the one with the greatest tidal influence on a galaxy. We define $\tht$ using massive galaxies as possible MD instead of all galaxies for two reasons, namely to be consistent with our earlier definition of $\dhst$ and also because surveys, including SDSS, provide high-completeness samples of massive galaxies down to this limit. Note that here we use the total dynamical mass $\mathcal{M}_{{\rm dyn}}$ for all the galaxies while $d_{jk}$ is the 3D distance between the pair of dwarf and massive galaxies. 

\section{Results} \label{sec:results}

In the previous section, we identified the sample of dwarfs with stellar masses $7.5 <{\rm log}(\mathcal{M}_{\ast}/M_{\odot})<9.5$ that are situated in low-mass halos with virial masses $9 < {\rm log}(\mathcal{M}_{200}/M_{\odot}) < 11.5$ as the field dwarf sample. In this section, we compare the quenched fractions of this sample with its satellite counterparts residing in massive halos (${\rm log}(\mathcal{M}_{200}/M_{\odot}) > 11.5$) and further study the characteristics of this sample. Thereby, we identify how their star-formation histories and baryonic content depend on their local and large-scale environments.

%and the effect of galactic conformity.

\subsection{Baryonic Content} \label{sec:baryoncontent}

\begin{figure*}
\centering
\includegraphics[scale=0.275]{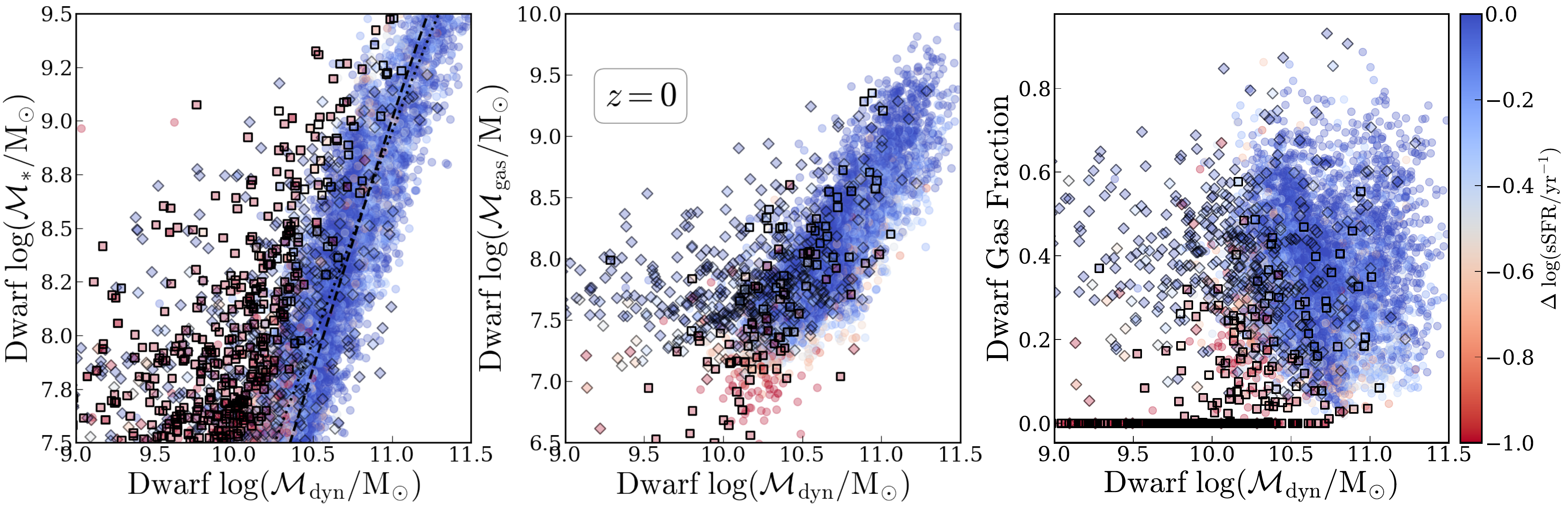}
\caption{ \textit{Left:} The stellar mass $\mathcal{M}_{\star}$ versus the total mass $\mathcal{M}_{\rm dyn}$ for the field dwarfs. The primaries (\textit{circles}), secondaries (\textit{diamonds}) and backsplash dwarfs (\textit{closed squares}) are colored according to their offset from SFMS. The black dashed line shows the best-fit stellar-halo mass relation (SHMR) from \citet{2013MNRAS.428.3121M}, whereas the black dotted line is the best-fit SHMR from \citet{2019MNRAS.488.3143B}. \textit{Center:} The gas mass $\mathcal{M}_{\rm gas}$ versus the total mass $\mathcal{M}_{\rm dyn}$. \textit{Right:} The gas fraction $f_{\rm gas}$ versus the total mass $\mathcal{M}_{\rm dyn}$. The backsplash and especially the secondary dwarfs are deficient in  $\mathcal{M}_{\rm dyn}$ at fixed $\mathcal{M}_{\ast}$, yet the latter remain gas-rich and star-forming. \label{fig:subhalo_mass}}
\end{figure*}

Before we proceed to the study of star-formation histories and environments of our field dwarf sample, we study the baryonic contents, i.e., stars and gas, of the galaxies at $z=0$. These properties are the primary observables for galaxies and we define quantities like the halo mass and the quenching timescale from them. First, we study how the stellar mass $\mathcal{M}_{\star}$ and gas mass $\mathcal{M}_{\rm gas}$ scale with respect to the total dynamical mass $\mathcal{M}_{\rm dyn}$ of the dwarf galaxies. In Fig. \ref{fig:subhalo_mass}, we show scatter plots with $\mathcal{M}_{\rm dyn}$ on all horizontal axes. All the points are colored by the offset from the SFMS and normalized with respect to this offset in the range of $-1<\Delta\ {\rm log}(sSFR/{\rm yr}^{-1})<0$. The bluer points are all star-forming dwarfs and the redder points show those dwarfs which are close to being quenched or are quenched already. The backsplash dwarfs are the square points with the edges, while the secondaries are the diamond points with thin edges. The rest of the primary dwarfs in the sample are depicted with the circular points. 

In the leftmost panel, we plot the stellar mass $\mathcal{M}_{\star}$ of the galaxies versus their total dynamical mass $\mathcal{M}_{\rm dyn}$. We also show with a black dashed line showing the best-fit stellar-halo mass relation (SHMR) from \citet{2013MNRAS.428.3121M} and the best-fit SHMR from \citet{2019MNRAS.488.3143B} with the black dotted line. We find that the primaries are mainly star-forming and situated along the central sequence of the SHMR. However, the backsplash and secondary dwarfs digress from this trend. The backsplash dwarfs are not only quenched, but are offset to the left of the primaries. Meanwhile, the secondaries are star-forming and are often offset further left of the central sequence. Alternately, at fixed $\mathcal{M}_{\star}$, the ${\rm log}(\mathcal{M}_{\rm dyn}/M_{\odot})$ of the backsplash dwarfs is $\lesssim 0.5$ dex with respect to the primaries. The lower dynamical mass is due to tidal stripping \citep{2021MNRAS.500.3957E} that removes dark matter from the outer regions of the galaxy. This is present for the secondaries as well, which have been plausibly tidally stripped but are still forming stars. This is notable because it shows that tidal stripping, which affects the outer halo of galaxies, has left the star-forming gas at its center intact. We also notice the presence of certain primaries that aren't backsplash galaxies but have been tidally stripped through past interaction(s) with a massive galaxy(s).

In the middle panel, we plot the dwarf gas mass $\mathcal{M}_{\rm gas}$ against the total dynamical mass $\mathcal{M}_{\rm dyn}$. Once again, we find that the star-forming primaries occupy a sequence, whereas the secondaries and other quenched primaries and backsplash dwarfs are offset from it. Below ${\rm  log}(\mathcal{M}_{\rm gas}/M_{\odot})\sim 7.5$, the galaxies are found to be quenched or are forming stars at moderately low rates. Here we note that some of the backsplash systems still retain their gas and are forming stars. Most of the secondaries, on the other hand, are offset from the primaries due to the effect of tidal stripping on $\mathcal{M}_{\rm dyn}$ that we found earlier. The effects of significant dark matter stripping without gas stripping for the secondary population are highlighted further in the right panel, where we plot the gas fraction defined as $f_{\rm gas} = \mathcal{M}_{\rm gas}/(\mathcal{M}_{\rm gas} + \mathcal{M}_{\star})$, against the dwarf mass $\mathcal{M}_{\rm dyn}$. An interesting observation that we make here is that 81\% of the backsplash galaxies have $f_{\rm gas}<0.1$, while for the primaries and secondaries this fraction is 7\% and 3\%, respectively. Out of these backsplash dwarfs, 96\% of them are quenched. This shows that the mass of the host halo that a dwarf interacts with significantly affects it evolution, because the backsplash dwarfs interact with the massive hosts in the denser environments, unlike their primary and secondary counterparts. 

%In the field, gas fractions of 0.8 are typical for dwarfs in this mass range \citep{2015ApJ...809..146B}.  

\subsection{Star-formation \& Stellar Mass} \label{sec:quench_frac_mstar}

Here we contrast the field and satellite samples to first resolve any dependence of their star formation on their stellar mass and also determine how much quenching is driven by the local and large-scale environments. These samples correspond to dwarfs residing in the SUBFIND-assigned FoF hosts with masses in the ranges $9 < {\rm log}(\mathcal{M}_{200}/ M_{\odot})< 11.5$ and ${\rm log}(\mathcal{M}_{200}/ M_{\odot})>11.5$, respectively. We define a galaxy as quenched if its sSFR is 1 dex below the SFMS (see Sec. \ref{sec:sfr_quench}). The quenched fractions henceforth in this work are evaluated as the ratio of the number of quenched galaxies with respect to the total number of galaxies in a particular bin. We evaluate and compare the quenched fraction in bins of stellar mass for the dwarf galaxies in both the field and satellite samples. To estimate the uncertainty, we perform $N=1000$ bootstrap resamplings of the host halos to generate distributions from which we evaluate the mean and the standard deviation. The lower panel of Figure \ref{fig:sfnms_geha12} shows these estimates in bins of ${\rm log}(\mathcal{M}_{\ast}/M_{\odot})$ for the total sample in black, the satellite and field dwarf samples in red and teal respectively. The shaded regions represent the bootstrap uncertainties. 

For the complete sample, the quenched fraction decreases from about 40\% to 20\% with increasing $\mathcal{M}_{\ast}$ of the dwarf galaxy, as has been shown in studies for satellite dwarfs around MW and analogous hosts \citep{2015ApJ...808L..27W,2022ApJ...933...47C,2024arXiv240414499G}. When we divide the sample by halo mass, we find that the field, or the low-mass host halo, sample is predominantly star forming, with quenched fraction at $< 20\%$ and a very weak dependence on $\mathcal{M}_{\ast}$.  This is unlike the satellite, or the high host mass halo, sample, for which the quenched fraction strongly varies with $\mathcal{M}_{\ast}$, from 75\% at the low-mass end to 45\% at the high-mass end. The field dwarf galaxies favor star-formation due to the fact that their halos are isolated from the environmental quenching processes prevalent around the massive galaxies. It is apparent that the dwarfs' local environment, i.e., membership with respect to their host halos, plays a significant role in whether a galaxy is still forming stars or not. In the subsequent sections, we explore whether the larger-scale environment plays an additional role.

%This result favors the halo-quenching model of \citet{2013MNRAS.428.3306W} which is the host halo mass as the primary determinant of quenching over the stellar mass even at the scales of $10< {\rm log}(\mathcal{M}_{200}/ M_{\odot})< 11.5$.

\subsection{Star-formation \& Environment} \label{sec:quench_frac}

\begin{figure*}
\centering
\includegraphics[scale=0.35]{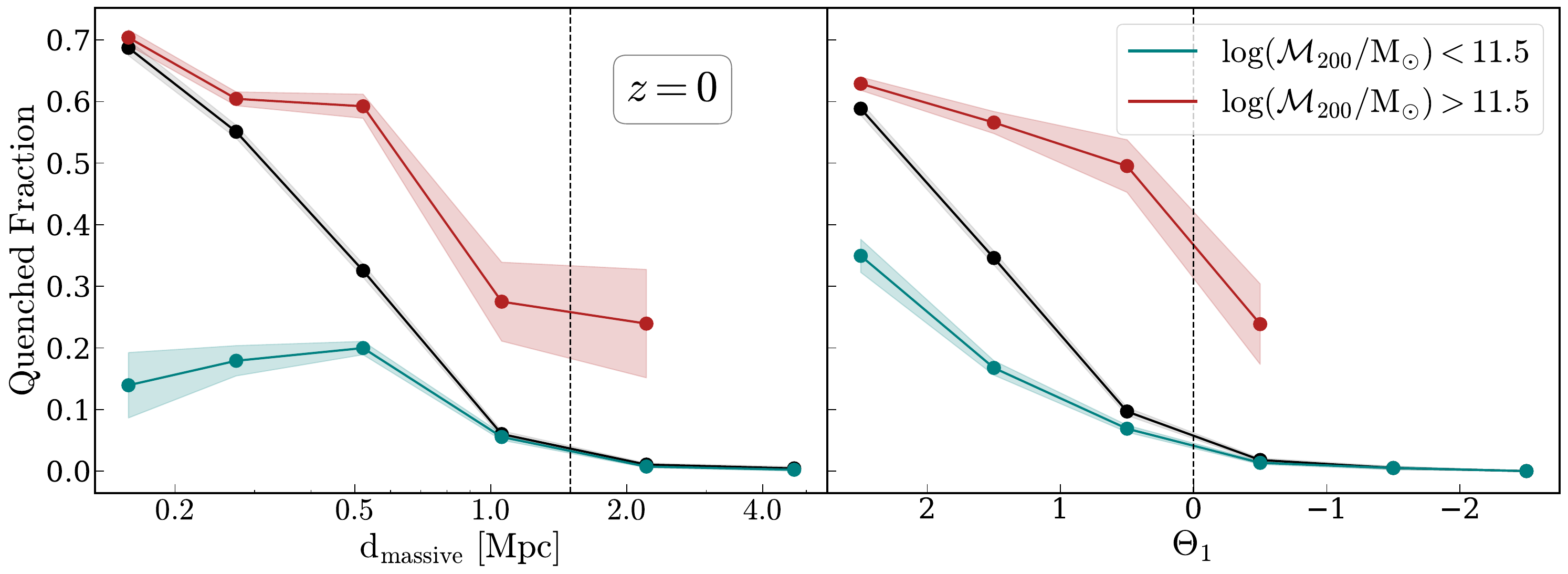}
\caption{ The quenched fraction for all the dwarf galaxies in TNG50 box, the field, and the satellite samples, in bins of the large-scale environmental density estimators, are plotted using the \textit{black, teal} and \textit{red} lines, respectively. The horizontal axes in the \textit{left panel} represents the distance to the nearest massive galaxy $\dhst$, while the \textit{right panel} shows the tidal index $\tht$.  All the axes have been orientated such that the large-scale environmental density decreases from left to right. The shaded regions represent uncertainties that are estimated using bootstrap resampling. The \textit{dashed} vertical lines show the limits $\dhst = 1.5$ Mpc and $\tht=0 $, respectively. We find that the quenched fraction depends on both the halo-scale and large-scale environments of the dwarf galaxies. \label{fig:quench_tracer}}
\end{figure*}

Here, we study the extremes of star-formation across different large-scale environments for the field dwarfs, and contrast this to the satellite dwarf sample. By extremes of star-formation, we namely look at the quenched fraction which we defined earlier in Sec. \ref{sec:sfr_quench} as those dwarfs with $\Delta\ {\rm log}(sSFR/{\rm yr}^{-1})<-1$, and also the starburst fraction, which we define as dwarfs with $\Delta\ {\rm log}(sSFR/{\rm yr}^{-1})>1$. We count these systems in bins of the two large-scale environmental density estimators $\dhst$ and $\tht$ in order to evaluate the quenched and starburst fractions, respectively. For $\dhst$, we define logarithmically spaced bins between 0.15 Mpc and 15 Mpc, while for $\tht$ we define linearly spaced bins between 3 and -3. This choice of binning has been done to accommodate the large dynamic range of $\dhst$. In Fig. \ref{fig:quench_tracer}, we plot the quenched fractions for the dwarf subhalos in bins of $\dhst$ and $\tht$ in the left and right panels, respectively. The quenched fractions and their uncertainties depicted here are derived from $N=1000$ bootstrap samples among the field dwarf host halos in consideration. 

The black points show the quenched fractions for all the dwarf galaxies in the simulation volume. For this sample, we find an environmental dependence on quenching wherein the dwarfs that are in proximity to a massive galaxy or in a tidally dense environment have a higher probability of being quenched, with a quenched fraction up to $\sim 70\%$. This shows the effect of large-scale environments in quenching dwarfs with stellar masses $7.5< {\rm log}(\mathcal{M}_{\ast}/M_{\odot})<9.5$. In similar fashion to Fig. \ref{fig:subhalo_mass}, the red and teal points represent the satellite and field dwarfs, respectively. The quenched fraction of the field dwarfs is sensitive to their large-scale environment at $\dhst<1.5$ Mpc and $\tht>0$, but is still reduced compared to the satellites.

It is evident that, at fixed large-scale environmental density represented by $\dhst$ and $\tht$, the quenched fraction is lower among the field dwarfs than the satellites. This establishes the role of the local environment on the dwarfs' star-formation.  The dwarfs residing in low-mass host halos have a lower quenched fraction, and vice versa. We consider this as a continuation of the trend in Fig. \ref{fig:sfnms_geha12}, where this divergence between the quenched fractions of field and satellite dwarfs was present at fixed stellar mass. Therefore, both the host halo (the local environment) and the large-scale environments of the dwarf galaxies significantly determine their star-formation rates, as opposed to their stellar masses.

\begin{figure*}
\centering
\includegraphics[scale=0.325]{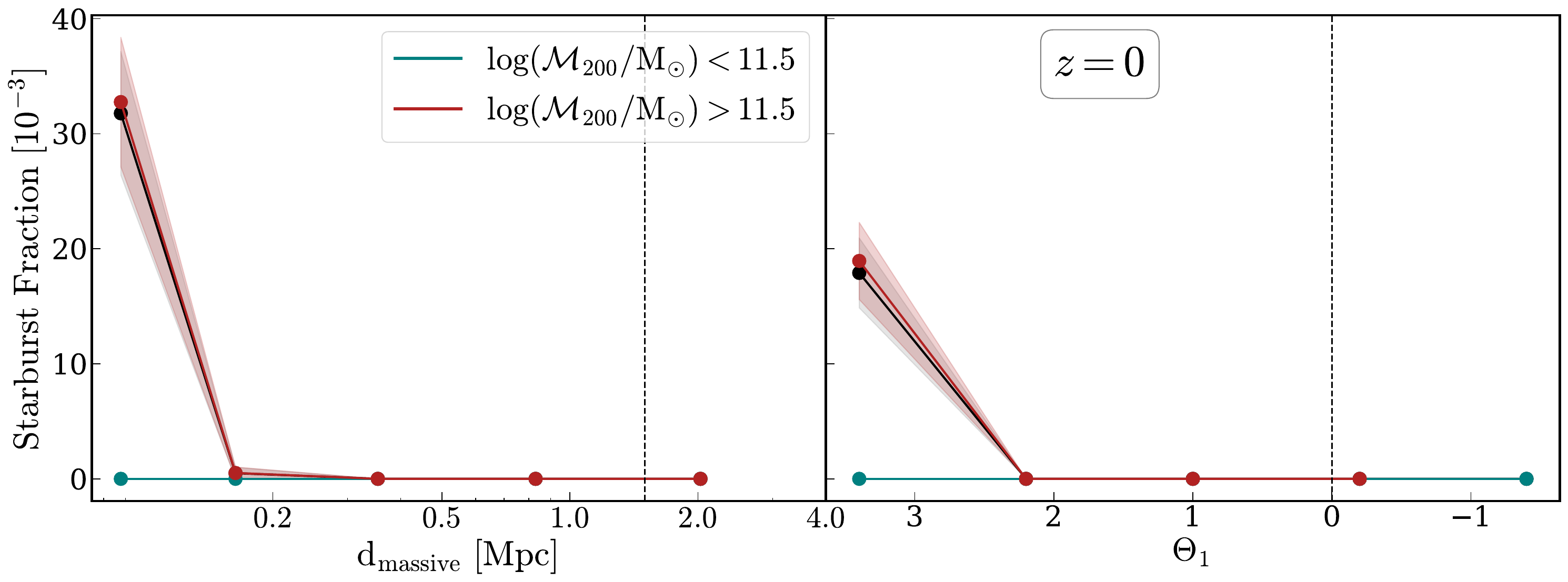}
\caption{ The starburst fractions for all the dwarf galaxies in TNG50 box, the field, and the satellite samples in bins of the large-scale environmental density estimators are plotted with the \textit{black, teal} and \textit{red} lines, respectively. The horizontal axes in the \textit{left panel} is the distance to the nearest massive galaxy $\dhst$, while the \textit{right panel} shows the tidal index $\tht$.  All the axes have been orientated such that the density of the large-scale environment decreases from left to right. The shaded regions represent uncertainties that are estimated using bootstrap resampling. The \textit{dashed} vertical lines show the limits $\dhst = 1.5$ Mpc and $\tht=0$, respectively. While starbursts are completely absent in the field dwarf sample, they are only present in the densest large-scale environment for the satellite dwarf sample. \label{fig:starburst_tracer}}
\end{figure*}

By contrast, the local environment plays the leading role in the starburst fraction, as we show in Fig. \ref{fig:starburst_tracer}. We plot the starburst fractions for the dwarf galaxies in bins of massive galaxy proximity $\dhst$ and tidal index $\tht$ in the left and right panels, respectively. Here we find that starbursts at the threshold of $\Delta\ {\rm log}(sSFR/{\rm yr}^{-1})>1$ are restricted to satellite dwarfs, and that there are no starbursts among the field dwarfs. The starburst fraction is $\sim 0.3\%$ in the bins closest to the massive hosts of the satellite dwarfs. This shows that the environmental processes necessary for driving an LMC/SMC analog into a starburst are absent in the low-density environment away from the massive galaxies. We already saw from the sSFR-$\mathcal{M}_{\ast}$ plot in the central panel of Fig. \ref{fig:sfnms_geha12}
 that most of our field dwarf sample lies along the SFMS and within $\pm 1$ dex of it. This further shows that the field dwarfs are isolated from the environmental processes of the massive galaxies that drive the extremes of states of star-formation,  either being quenched or experiencing a starburst.
 
%In order to select the groups of our interest we search for halos in which the most massive subhalo has a stellar mass in the range of 10$^8 M_{\odot} < \mathcal{M}_{\star} <$ 10$^{11} M_{\odot}$. While the most massive subhalo corresponds to the central galaxy in the group, the rest of the subhalos of lesser mass make up the satellite galaxies in the group.

\begin{figure*}
\centering
\includegraphics[scale=0.3]{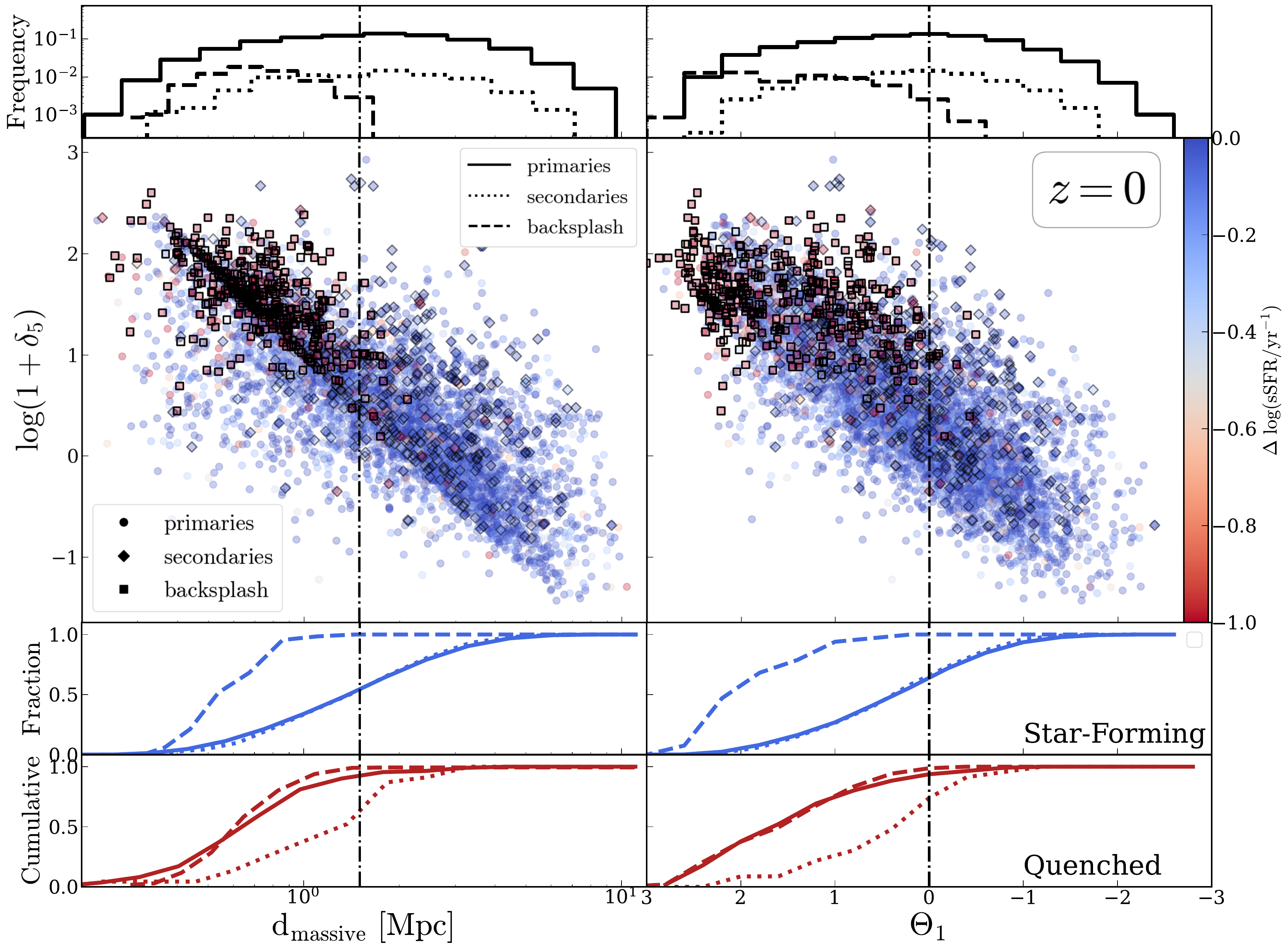}
\caption{The distributions of the field dwarf sample in $\dhst-{\rm log}(1+\delta_5)$ (\textit{left}) and $\tht -{\rm log}(1+\delta_5)$ (\textit{right}) spaces. ${\rm log}(1+\delta_5)$ is the fiducial density estimator (the overdensity parameter) and we find that this sample corresponds to ${\rm log}(1+\delta_5)\lesssim2$. The primary dwarfs (\textit{circles}) and secondary dwarfs (\textit{diamonds}) and backsplash dwarfs (\textit{closed sqaures}) are colored according to their sSFR deviation from the SFMS, with bluer colors indicating active star-formation as opposed to saturated red color implying a quenched state. The backlsplash dwarfs are found in relatively high density environments ($\dhst<1.5 {\rm Mpc};\tht>0$) and are quenched while the primaries and secondaries are more isolated and show ongoing star-formation. In the \textit{lower} panel we plot the cumulative fractions of star-forming and quenched field dwarfs, respectively, with \textit{blue} and \textit{red} lines, respectively. 
\label{fig:env_dssfr}}
\end{figure*}

We continue our investigation into environment and star-formation in the field dwarfs by focusing on the field dwarfs and its three sub-samples. In Fig. \ref{fig:env_dssfr}, we analyze the distributions of our field sample in the density estimator spaces where we have the overdensity parameter ${\rm log}(1+\delta_5)$ along the vertical axes and $\dhst$ and $\tht$ along the horizontal axes of the left and right columns, respectively. Each point in the distributions is colored according to the deviation of sSFR from the SFMS in the range of $-1<\Delta({\rm log sSFR})<0$ to visually depict the gradient in sSFR between the SFMS and the quenching threshold, corresponding to the blue and red ends of the colormap normalization. At the top, we show the histograms for the distributions of $\dhst$ and $\tht$ for the three classes of field dwarfs in this work --- primaries, secondaries and backsplash. In the lower panel, we plot the cumulative fractions for the star-forming and quenched field dwarfs in blue and red, respectively. 

We showed earlier, in the right panel of Fig. \ref{fig:dens_map}, that field dwarfs occupy the range $-2 \lesssim {\rm log}(1+\delta_5) \lesssim 2$. From the two panels of Fig. \ref{fig:env_dssfr}, we find that this compares to a range 0.2 Mpc $<\dhst<$ 10 Mpc and $3.00>\tht>-2.39$, with median values of 1.69 Mpc and 0.25, respectively. About 92\% of the field dwarfs are star-forming and 8\% are quenched. The quenched backsplash and primary dwarfs mainly occupy overdensities of $1 \lesssim {\rm log}(1+\delta_5) \lesssim 2$ and constitutes the feature in the lower-right panel of Fig. \ref{fig:dens_map} corresponding to host masses ${\rm log}(\mathcal{M}_{200}/M_{\odot})\lesssim 10$. This roughly corresponds to the $\dhst<1.5$ Mpc and $\tht>0$ regions of the estimator spaces. According to the cumulative fractions, these regions contain 91\% and 93\% of the quenched dwarfs in the sample, respectively. On the other hand, the field dwarf selections with $\dhst>1.5$ Mpc and $\tht<0$ are predominantly star-forming primaries, with only $\sim 1\% $ being quenched in both cases. These estimates are robust to the definition of the field, as we discover in Appendix \ref{sec:outside_halo}. Nonetheless, we find that by defining the dwarfs outside the virial volume of the massive hosts as the field sample (as in Appendix \ref{sec:outside_halo}, and as opposed to our FoF group-based definition for our fiducial field sample) leads to greater diversity in the density estimators for $\dhst<1.5$ Mpc and $\tht>0$.

Nonetheless, it stands out that while the field dwarfs are in majority star-forming and shaped by their local environment, the gradient in the large-scale environmental density has a non-negligible effect on star formation. This is best seen in the uppermost and lowermost panels of Fig. \ref{fig:env_dssfr}. The distributions of the three sub-samples and the cumulative distributions of the star-forming and quenched dwarfs are shown here. The backsplash and secondary dwarfs have similar distributions in $\dhst$ and $\tht$ regardless of their star-formation. On the other hand, the primary dwarfs that are star-forming and quenched have distributions similar to the secondary and backsplash dwarfs, respectively. This shows that while the overall populations of primaries and secondaries are similar in terms of star-formation, the subset of quenched primaries have properties that are similar to the backsplash dwarfs in the regions with $\dhst<1.5$ Mpc and $\tht>0$.

\subsection{Star-formation Histories} \label{sec:SFH}

\begin{figure*}
\centering
\includegraphics[scale=0.325]{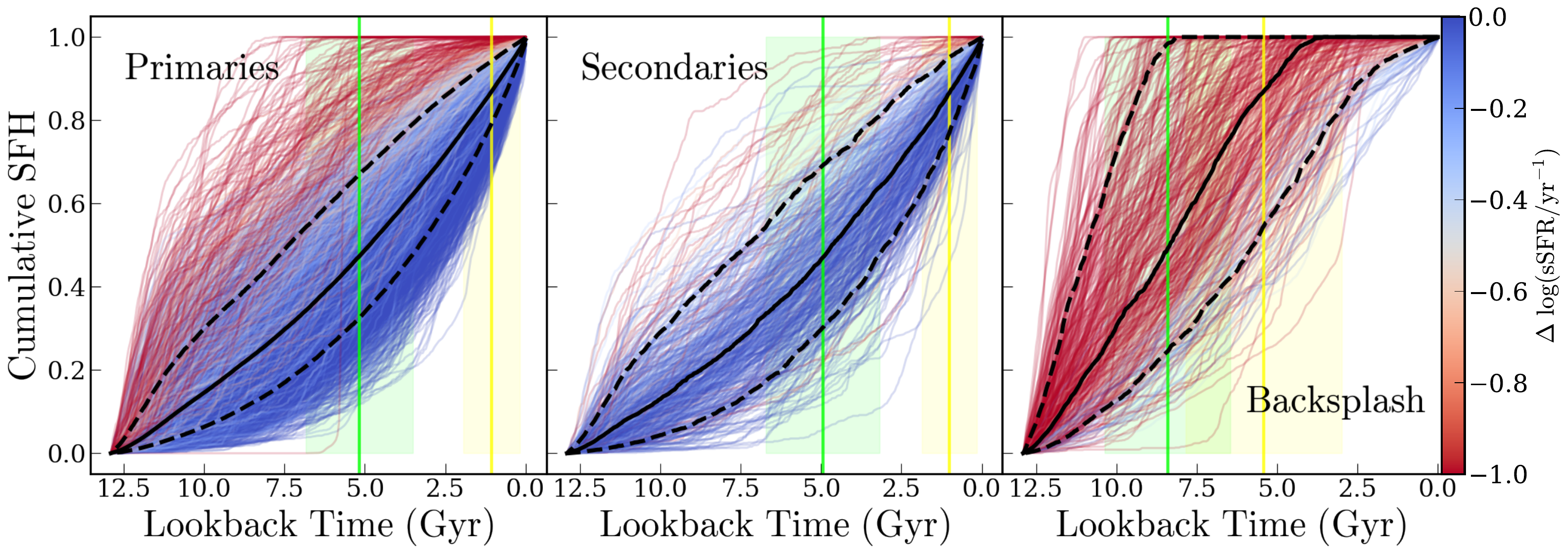}
\includegraphics[scale=0.35]{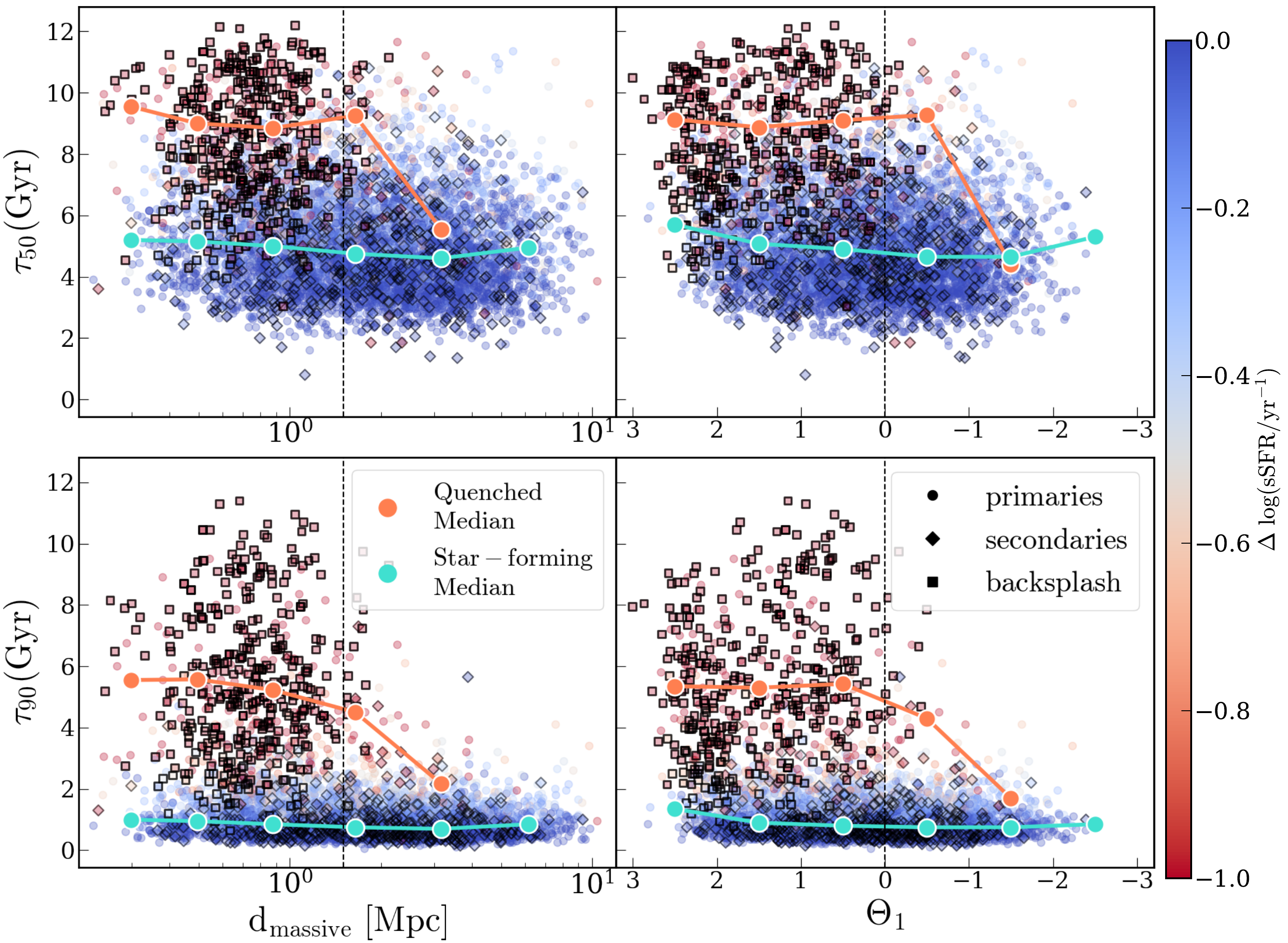}

\caption{The cumulative star-formation histories (SFH) of the three sub-samples of field dwarfs --- primaries (\textit{top left}), secondaries (\textit{top center}) and backsplash dwarfs (\textit{top right}). The solid lines are median of the distributions whereas the upper and lower black dashed lines represent the 25$^{\rm th}$ and 75$^{\rm th}$ percentiles, respectively. The mean and standard deviation of the epochs $\tau_{50}$ and $\tau_{90}$ in units of Gyr for each sub-sample (see Table \ref{tab:t50_t90})) are plotted with the \textit{green} and \textit{yellow vertical bands} respectively. The epochs $\tau_{50}$ (\textit{center row}) and $\tau_{90}$ (\textit{bottom row}) for the full sample of field dwarfs are plotted against their density estimators $\dhst$ (\textit{left column})and $\tht$ (\textit{right column}). The primary dwarfs (\textit{circles}) and secondary dwarfs (\textit{diamonds}) and backsplash dwarfs (\textit{closed sqaures}) are colored according to their offset from the SFMS. The \textit{red} and \textit{blue} profiles show the medians of $\tau_{50}$ and $\tau_{90}$ for the quenched and star-forming dwarfs in bins of the estimators. The backsplash dwarfs have larger $\tau_{50}$ and $\tau_{90}$ by virtue of them being quenched earlier.
\label{fig:cumulativeSFH}}
\end{figure*}

\begin{table}
    \centering
    \begin{tabular}{|c|c|c|c|}
    \hline
Class & QF (\%) & $\tau_{50}$ (Gyr) & $\tau_{90}$ (Gyr) \\
\hline
Primaries & 2.2 & 5.2 $\pm$ 1.7 & 1.1 $\pm$ 0.9 \\
Secondaries & 4.9 & 4.9 $\pm$ 1.8 & 1.0 $\pm$ 0.9 \\
Backsplash & 82.4 & 8.4 $\pm$ 2.0 & 5.4 $\pm$ 2.4 \\
\hline
Star-forming & 0 & 5.1 $\pm$ 1.6  & 1.0 $\pm$ 1.8 \\
Quenched  & 100 & 8.8 $\pm$ 1.9  & 5.6 $\pm$ 1.9 \\
\hline
    \end{tabular}
    \caption{ Quenched fractions and the mean and standard deviation estimates for the star-formation epochs $\tau_{50}$ and $\tau_{90}$ for the different classes of field dwarf subhalos. }
    \label{tab:t50_t90}
\end{table}

We study the star-formation histories of the three sub-samples of dwarfs and study its variation across large-scale environment. We compute the cumulative star-formation histories (SFH) by following the formalism of \citet{2021MNRAS.508.1652J}. We take the scale-factor of when the star particle was formed, convert this to a lookback time, and compute the histogram of this time weighted by the masses of the star particles. From the SFHs, we also derive the lookback times by which these dwarf galaxies assembled 50\% and 90\% of their present day stellar mass, which are $\tau_{50}$ and $\tau_{90}$ respectively \citep{2015ApJ...804..136W,2021MNRAS.508.1652J,2024ApJ...961..236C}.  In Table \ref{tab:t50_t90}, we list the summary statistics for these estimates for each field dwarf sub-sample as well as those that are star-forming and quenched. 

The cumulative normalized histograms indicative of the SFH for the three classes of field dwarfs are plotted as a function for lookback times in the topmost panel of Fig. \ref{fig:cumulativeSFH}, where each line is coded by the offset from SFMS in the range of $-1<\Delta\ {\rm log}(sSFR/{\rm yr}^{-1})<0$. The solid black lines are medians of the curves, whereas the upper and lower black dashed lines represent the 25$^{\rm th}$ and 75$^{\rm th}$ percentiles of the curves, respectively. In the same panel, we plot the mean and standard deviation of the epochs $\tau_{50}$ and $\tau_{90}$ for each sub-sample of field dwarfs with the green and yellow vertical bands, respectively. We explore the dependence of these timescales on the large-scale environment by plotting $\tau_{50}$ and $\tau_{90}$ against density estimators $\dhst$ and $\tht$ in the center and lowermost panels of Fig. \ref{fig:cumulativeSFH}. Once again, the points are colored according to the sSFR offset from the SFMS. In the same plots, we also show with the red and blue lines the medians of $\tau_{50}$ and $\tau_{90}$ for the quenched and star-forming dwarfs that are evaluated in bins of the estimators. We find that there is little environmental dependence on the medians of $\tau_{50}$ and $\tau_{90}$ for the star-forming dwarfs and the quenched dwarfs residing in the denser region. However around $\dhst \sim 1.5$ Mpc and $\tht \sim 0$, the quenched dwarf abundance diminishes and with it their medians approach those of their star-forming counterparts, which shows the large-scale environment modulating the quenched dwarfs' SFHs at these scales.

We find by inspecting the cumulative SFHs that the primaries and secondaries are mostly star-forming with comparable estimates of $\tau_{50}\sim 5$ Gyr and $\tau_{90}\sim 1$ Gyr, and quenched fractions of 2.2\% and 4.9\% respectively. The quenched dwarfs that make up 8\% of the field sample mainly dwell in $\dhst<1.5$ Mpc and $\tht>0$, and show earlier epochs of $\tau_{50}\sim 8$ Gyr and $\tau_{90}\sim 5$ Gyr. The quenched dwarfs are mainly made up of the backsplash and some primary dwarfs, with the backsplash dwarfs being largely quiescent with an overall 84\% quenched fraction. Although both secondaries and backsplash dwarfs have been heavily tidally stripped (see Sec. \ref{sec:baryoncontent}), it is interesting to note that only the latter are quenched while the secondaries are mainly star-forming.

The median $\tau_{50}$ and $\tau_{90}$ is found to be independent on the large-scale environment for the star-forming dwarfs whereas for the quenched dwarfs there is a weak dependence. This is connected to the fact that the 92\% of the overall field sample is star-forming there is a higher quenched fraction in the denser environments ($\dhst<1.5$ Mpc and $\tht>0$). Here we also find a greater dispersion of $\tau_{90}$ compared to that of $\tau_{50}$ between the quenched and star-forming populations here. This follows our results from Sec. \ref{sec:quench_frac} and \ref{sec:quench_frac_mstar}, that the quenched dwarfs that live in the denser environment have been processed by their large-scale environment. The median epochs for the quenched dwarfs approach those of their star-forming dwarfs at $\dhst>1.5$ Mpc and $\tht<0$. Our result is relevant to Fig. 14 of \citet{2024ApJ...961..236C} where the mean $\tau_{90}$ is seen to increase with decreasing distance to a massive neighbor. However their result pertains to dwarfs with masses down to ${\rm log}(\mathcal{M}_{\ast}/M_{\odot})=6$, in the regime where they are quenched more often and earlier. Furthermore, their trend arises from all the dwarfs, whereas we separate them as quenched and star-forming. These caveats aside we note that the quenched fraction and star-formation timescales of the field dwarfs in both studies show a weak dependence on large-scale environment.

%The median epochs for the quenched sub-population are found to have a weak dependence on the large-scale environment at $\dhst \sim 1.5$ Mpc and $\tht \sim 0$).

\subsection{Galactic Conformity} \label{sec:galactic_conform}

\begin{figure*}
\centering
\includegraphics[scale=0.35]{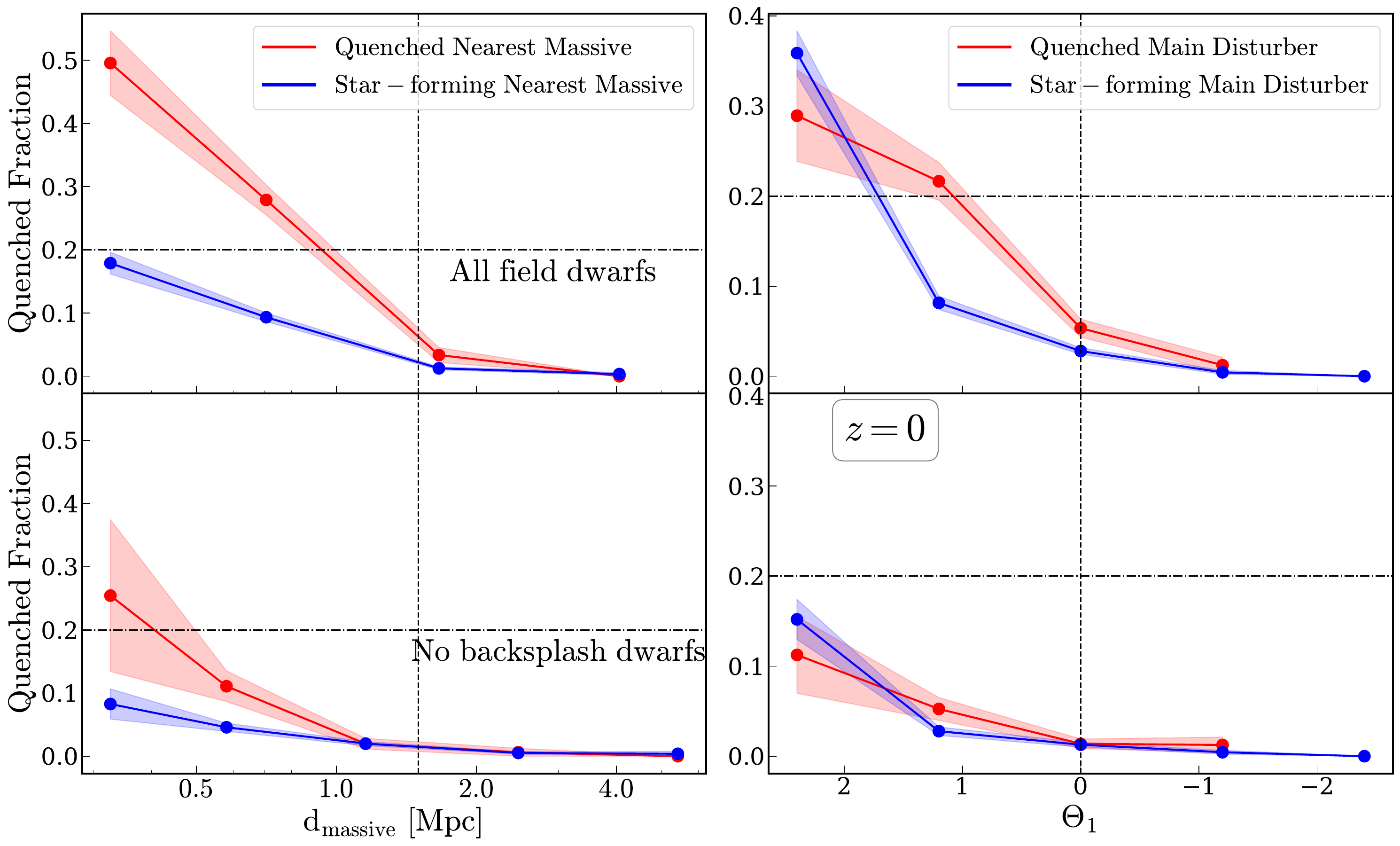}
\caption{The quenched fraction for the field dwarf sample simulated in the TNG50 box as a function of the density estimators ---  \textit{left:} the distance to the nearest massive galaxy $\dhst$,  \textit{right:} the tidal index $\tht$. All the axes have been orientated such that the density of the large-scale environment decreases from left to right. The red and blue profiles  represent the field dwarfs selected around quenched and star-forming massive galaxies, respectively. While in the \textit{upper left} panel we find the effect of two-halo galactic conformity, when we remove the backsplash dwarfs from our sample the difference in the quenched fractions is removed in the \textit{lower left} panel.  The \textit{dashed} vertical lines show the limits $\dhst = 1.5$ Mpc and $\tht=0$, respectively. \label{fig:quench_conform}}
\end{figure*}

The preceding work aimed at studying the connection between star-formation in field dwarfs and their large-scale environment, directs us to investigate the presence of two-halo conformity in this sample. This will be measurable as a higher quenched fraction for the field dwarfs located in proximity to the quenched massive galaxies \citep{2013MNRAS.430.1447K,2023ApJ...943...30O}. Following the exercise in Sec. \ref{sec:quench_frac}, we plot the quenched fraction in bins of the density estimators $\dhst$ or $\tht$ for the field dwarfs, for two cases based on the state of the proximal massive galaxy as either star-forming or quenched. The massive galaxy is the nearest neighbor and the Main Disturber that is used to determine the $\dhst$ or $\tht$ correspondingly. The result is shown in the upper panel of Fig. \ref{fig:quench_conform}, with the red and blue points along with their shaded regions being bootstrap estimates of the quenched fraction of field dwarfs around the quenched and star-forming massive galaxies, respectively. Noticeably for the $\dhst$ estimator, the quenched fraction around the quenched massive host is larger at small $\dhst$ and significant given the uncertainties. This shows the strong clustering of quenched galaxies that is a signature of the two-halo galactic conformity. For the $\tht$ estimator, the same effect is weaker but still present. 

We investigate this effect further by removing the backsplash dwarfs from the field dwarf sample, and again measure the quenched fraction as functions of the density estimators. We find that the two-halo galactic conformity to be diminished and not significant given the uncertainties, especially if we characterize the large-scale environment by $\dhst$. For both estimators, the magnitude of the quenched fraction is reduced by $\sim 50-70\%$ upon removing the contribution from the backsplash dwarfs, but still increases for small $\dhst$ and large $\tht$. These results, while highlighting the importance of the two-halo galactic conformity at the scales of $\dhst \lesssim 1$ Mpc, also show that they are mainly driven by the backsplash dwarfs. However, additional environmental processes are needed to explain the non-zero quenched fraction of the non-backsplash field dwarfs at these scales. In Appendix \ref{sec:outside_halo}, we find that the two-halo galactic conformity and its relation with the backsplash galaxies, persists even with a broader sample of field dwarfs, showing the robustness of our result.

\section{Discussion} \label{sec:discussion}

In this study, we have investigated the low quenched fractions of LMC/SMC analogs,  particularly looking at the role played by their host halos and their large-scale environment. We isolate a sample of dwarf galaxies in the field of the TNG50 simulation, characterize their environment and star-formation, and study the connection between the two. In this section, we highlight our results and relate these with the contemporary understanding of such systems in the literature.

\subsection{Field Dwarfs and their Large-scale Environments}

To understand the large-scale environments of the halos hosting dwarfs with stellar masses $7.5 <{\rm log}(\mathcal{M}_{\ast}/M_{\odot})<9.5$ we find that dwarfs with hosts having, $9 < {\rm log}(\mathcal{M}_{200}/M_{\odot}) < 11.5$, are LMC/SMC analogs in the field of the TNG50 simulation volume as shown by the values of the overdensity parameter of ${\rm log}(1+\delta_5) \lesssim 2$ (see Fig. \ref{fig:dens_map}).  This sample still possesses a considerable $\sim 2$ dex diversity in ${\rm log}(1+\delta_5)$, ranging from the filaments to the voids of the simulation box. 

%Therefore, we further use two commonly employed measures of the large-scale environment of galaxies --- the tidal index $\tht$ and the distance to the nearest massive galaxies $\dhst$ --- as density estimators (see Fig. \ref{fig:env_dssfr}).}

%Among the low-mass dark matter halos at $z=0$, the ones in the high-density environments are satellites within a massive halo, whereas those in the low-density environments are most often centrals.

The biggest difference in star-formation properties is between the field dwarfs and their satellite counterparts inside hosts having, ${\rm log}(\mathcal{M}_{200}/M_{\odot}) > 11.5$. We find that the field dwarf sample is largely star-forming, with an overall quenched fraction of $8\%$ (see Fig. \ref{fig:sfnms_geha12} and \ref{fig:quench_tracer}). Intriguingly, we do not find starbursts in the field (see Fig. \ref{fig:starburst_tracer}). This is in contrast to the satellite dwarfs of massive galaxies, which have a higher quenched fraction of $\gtrsim 50\%$ and also account for all dwarf galaxies undergoing starbursts. 

These trends also depend strongly on stellar mass and the density estimators for the satellites, as opposed to their field counterparts. This shows that, compared to the satellite dwarfs in the denser environments, the field dwarfs have less diversity in their sSFR with respect to their low-density environment \citep{2024ApJ...961..236C}. This is seen in Fig. \ref{fig:sfnms_geha12}, where the field dwarfs essentially lie along the SFMS in the sSFR-$\mathcal{M}_{\ast}$ space with sSFR $\sim 0.1 \ {\rm Gyr}^{-1}$, whereas they are few in the extremes of star-formation, characterized by sSFR $> \mp 1$ dex of the SFMS. Quenched and starburst dwarfs are to be mainly found in the environment around massive galaxies in clusters and groups, with the quenched fractions increasing with decreasing distances toward their centers.

%We identify this by studying the quenched and starburst fractions corresponding to the fraction of dwarfs with $\Delta({\rm log sSFR})>\mp1$ respectively, in bins of their stellar mass and the density estimators. There is less diversity within the field population.

Previous authors have attempted to disentangle the causes of quenching as being tied to either halo-scale environment or stellar mass \citep{2010ApJ...721..193P,2014MNRAS.438..717K}. However, in this study, the selection of the field dwarfs is driven by their host mass $\mathcal{M}_{200}$, as seen in Fig. \ref{fig:dens_map}. This shows that while the halo mass is the primary driver \citep{2016MNRAS.457.4360Z,2018MNRAS.476.1637Z}, their large-scale environment matters too \citep{2012MNRAS.419.2670M,2012MNRAS.419.2133H,2013MNRAS.428.3306W} in modulating their star-formation.

%While the overall quenched fraction is low for dwarfs with stellar masses $7.5 <{\rm log}(\mathcal{M}_{\ast}/M_{\odot})<9.5$, they still have great diversity in star-formation that is driven by the diversity of their large-scale environments, as shown by comparing field dwarfs in observation \citep{2012ApJ...757...85G} with their satellite counterparts around MW or its analogs \citet[e.g.][]{2015ApJ...804..136W,2016MNRAS.455.4013D,2024arXiv240414499G}. 

Another crucial insight that we get is by placing the two density estimators on an equal footing and comparing them. While $\dhst$ is relatively easy to measure from the positions of neighboring galaxies, the measurement of $\tht$ explicitly requires the dynamical mass for the neighbors as well, which is calculated from their luminosities by assuming a mass-to-light ratio. Nonetheless, $\tht$ traces the mass distribution and thereby the potential of the large-scale matter distribution better than $\dhst$. This results in $\tht$ having greater contrast in the density variations in regions with $\tht>0$ and $\dhst<1.5$ Mpc. This is why, although the distributions in Fig. \ref{fig:env_dssfr}, \ref{fig:cumulativeSFH} and \ref{fig:quench_conform} look similar between $\dhst$ and $\tht$, they are not exactly the same.

This divergence aside, there is a notable similarity between these two estimators, as they well select samples of field galaxies that are isolated and star-forming. We find that the field dwarf sample that resides in low-density environments of $\dhst>1.5$ Mpc and $\tht<0$ both have an overall quenched fraction of about 1\%. This agreement is remarkable since the estimators have different motivations \citep{2012ApJ...757...85G,2013AJ....145..101K}. Nonetheless, we find that 8\% of the field dwarf sample is quenched. \citet{2021ApJ...915...53D} studied the quenched fractions of isolated galaxies down to ${\rm log}(\mathcal{M}_{\ast}/M_{\odot})=8$ further using the EAGLE, TNG100, and SIMBA simulations and found reasonable agreement with measurements from the SDSS survey.

Our quenched sample mainly consists of the backsplash dwarfs, along with some primaries and secondaries in dwarf associations. The large-scale environments corresponding to $\dhst<1.5$ Mpc and $\tht>0$ contain 88\% and 92\% of the quenched field dwarfs. This also corresponds to overdensities of $1 \lesssim {\rm log}(1+\delta_5) \lesssim 2$, and here the primaries are subject to the gravitational influence of massive galaxies, even though they are well outside their virial radius \citep{2013MNRAS.430.3017B,2021ApJ...909..112D,2023arXiv231200773B}. Such environments are greatly relevant for the survey of quenched LMC/SMC analogs. We continue the discussion on these objects in the following section.

%Geha used the 1.5 Mpc criterion, based empirically that the quenched fraction seemed to plateau beyond that.  You found the physical driver for it.  I'd recommend using this as a bridge to the next paragraph.

\subsection{Backsplash \& Quenched Primary Dwarfs}

The fraction of the field dwarfs that are quenched in the neighborhood of massive galaxies show some significant attributes that we discuss here. For example, in Fig. \ref{fig:subhalo_mass}, we find that the quenched backsplash dwarfs have stellar masses spanning the full range $7.5 <{\rm log}(\mathcal{M}_{\ast}/M_{\odot})<9.5$, while the quenched primaries are of lower mass $7.5 <{\rm log}(\mathcal{M}_{\ast}/M_{\odot})<8.0$. This connects with the finding of \citet{2014MNRAS.442.1396W,2015ApJ...804..136W,2015MNRAS.447..698P,2024arXiv240414499G} that galaxies with ${\rm log}(\mathcal{M}_{\ast}/M_{\odot})\gtrsim 8$ are difficult to quench environmentally. The scale of ${\rm log}(\mathcal{M}_{\ast}/M_{\odot})\sim 8$ is critical, as demonstrated by \citet{2018MNRAS.478..548S,2021ApJ...909..139A,2023MNRAS.519.4499P}, because dwarfs less massive than this are prone to be quenched faster by environmental processes. At the mass scales of our dwarf sample, AGN feedback is not significant in removing their gas
\citep{2021MNRAS.500.4004D}. Rather, stellar-feedback-driven outflows in the dwarfs are easier to ram-pressure strip than their interstellar medium \citep{2022arXiv220909262G}. The mass-loading factor rises dramatically as a function of decreasing stellar mass, which makes dwarfs with ${\rm log}(\mathcal{M}_{\ast}/M_{\odot})\lesssim 8$ easy to quench.

We also find that the halos of these dwarfs have been subject to tidal-stripping in the leftmost panel of Fig. \ref{fig:subhalo_mass}. For a more massive backsplash sample with $9.7 <{\rm log}(\mathcal{M}_{\ast}/M_{\odot})<10.5$, \citet{2014MNRAS.439.2687W} show that, on average, halo mass is reduced compared to non-backplash centrals, primarily due to tidal stripping inside the massive halo \citep{2011MNRAS.412..529K}. Furthermore, \citet{2023MNRAS.520..649B} show that this affects the density profiles of the backsplash dwarfs. Probing the weak-lensing profiles of these backsplash dwarfs is a way to investigate the extent of tidal stripping on their halos. This can ultimately be used to reduce the scatter in the SHMR at scales of ${\rm log}(\mathcal{M}_{\rm dyn}/M_{\odot})<10$ \citep{2021ApJ...923...35M,2024ApJ...961..236C}.

The neighboring massive galaxies of the quenched backsplash dwarfs in Fig. \ref{fig:quench_mneigh} are mostly located in the cluster-scale halos with ${\rm log}(\mathcal{M}_{200}/M_{\odot})\gtrsim 13$. This is complementary to the findings of \citet{2014MNRAS.439.2687W}, that backsplash galaxies with ${\rm log}(\mathcal{M}_{\ast}/M_{\odot})\sim 9$ make up $\sim 50\%$ of the centrals outside cluster-scale halos as massive as this. Upon closely inspecting the backsplash systems in Fig. \ref{fig:backspl_rads}, we find that the median lookback time for the last pericentric passages is 3.6 Gyr, which is later than the average $\tau_{90}= 5.4$ Gyr for the same systems. The dynamical times imply that the backsplash dwarfs have already experienced starvation and ram pressure \citep{2018MNRAS.478..548S,2022arXiv220909262G}, while on infall or previous pericentric passages \citep{2004PASJ...56...29F,2017MNRAS.467.3268R,2021MNRAS.500.4004D}. All these processes serve to deplete backsplash galaxies' gas reservoirs, as shown by 81\% of their population having $<10\%$ gas fractions in Fig. \ref{fig:subhalo_mass}.

The environments outside the virial radius of a massive halo can play a role in quenching  \citep{2013ApJ...763L..41B,2013MNRAS.430.3017B}, for example, the filaments in the cosmic web \citep{2019OJAp....2E...7A,2022MNRAS.tmp.3505P}. \citet{2023ApJ...950..114H} showed that stripping of cold gas out to $\sim 1$ Mpc from nodes and filaments in the cosmic-web contributes to the suppression of sSFR of dwarfs with $8 <{\rm log}(\mathcal{M}_{\ast}/M_{\odot})<9$ in TNG100. In a subsequent work \citet{2025arXiv250113159B} have also shown that in the same stellar mass range of TNG50, the quenched primaries are subject to stripping while crossing the filaments and sheets in the cosmic-web. These examples of quenched field dwarfs both in our and contemporary works follow the trend of brighter centrals \citep{2014MNRAS.444.2938H,2014MNRAS.439.2687W,2021MNRAS.505..492A}, that are environmentally quenched yet are found up to a few Mpc from massive galaxies.

This further manifests as the two-halo galactic conformity, wherein the measurements of clustering and star-formation rates of galaxies are found to be correlated. Observations and simulations \citep{2014MNRAS.439.2687W,2018MNRAS.477..935T,2022MNRAS.513.2271L,2023MNRAS.523.1268W,2023MNRAS.519.1913A} show that the conformity effect is found among massive galaxies in the field, and is a result of quenched backsplash and primary galaxies in the vicinity of massive host halos. In Sec. \ref{sec:galactic_conform}, we show that the backsplash dwarfs can explain most, if not all, of the two-halo galactic conformity effect for the relevant stellar mass scales. This is because the backsplash dwarfs are largely located near the cluster-scale halos with ${\rm log}(\mathcal{M}_{200}/M_{\odot})> 13$ that tend to be assembled earlier and also host quenched massive galaxies. A residual quenching signal still persists, and is likely from the primaries that have been quenched by other environmental processes \citep{2025arXiv250113159B}.

Our study also highlights that star-formation among the field dwarfs is modulated by the large-scale structure in the simulation rather than an individual halo of a massive galaxy. The most massive cluster in the TNG50 box has ${\rm log}(\mathcal{M}_{200}/M_{\odot})=14.26$, corresponding to which the the virial radius is $\mathcal{R}_{200} = 1.20$ Mpc, and the splashback radius \citep{2015ApJ...810...36M} is $\mathcal{R}_{sp} = 1.41$ Mpc \footnote{The splashback radius has been calculated using Eq. 7 of \citet{2015ApJ...810...36M} aided by the \texttt{COLOSSUS} package \citep{2018ApJS..239...35D}.} We also read from the right panel of Fig. \ref{fig:backspl_rads} that for this cluster, the backsplash dwarf with the largest cluster-centric distance is 2.26 Mpc. It is to be noted that this is larger than not only the virial radius, but also the splashback radius for the cluster. In all, we find that the large-scale environments of massive halos contribute to the isolation criteria $\dhst>1.5$ Mpc, as opposed to it being determined by their individual virial radii \citep{2014MNRAS.442.1396W}. This needs to be investigated further, but there is only a single cluster with ${\rm log}(\mathcal{M}_{200}/M_{\odot})>14$ in TNG50. With a larger simulation box at a finer mass resolution, a large number of backsplash galaxies may be sampled, and therefore the extent of a massive halo's influence on the isolation threshold can be investigated.

\subsection{Dwarf Associations}

Our study quantifies the incidence of secondaries alongside the primaries in the range of masses $7.5 <{\rm log}(\mathcal{M}_{\ast}/M_{\odot})<9.5$. Such groups of dwarfs that are gravitationally bound have been found in the local universe \citep{2002ApJ...569..573T,2016MNRAS.459.1827P,2017NatAs...1E..25S}. Of these we find that 317 are pairs, 45 are triples and 5 have more than three dwarfs. It should be noted that we are only able to resolve down to ${\rm log}(\mathcal{M}_{\ast}/M_{\odot}) = 7.5$. However, if we were to resolve below this limit, we expect to find more substructure in the dwarf groups. \citet{2017MNRAS.472.1060D} show that 1-6 companions with ${\rm log}(\mathcal{M}_{\ast}/M_{\odot})> 5$ for LMC sized dwarf can be expected. 

Understanding the evolution of dwarfs in associations in the field is of importance given their role in the assembly of massive galaxies \citep{2008ApJ...686L..61D,2023MNRAS.525.3849S}. For example, we understand that the LMC/SMC group has, after its infall, contributed to the satellite population of MW 
\citep{2015MNRAS.453.3568D,2020ApJ...893...48N,2020ApJ...893..121P,2020MNRAS.495.2554E,2022ApJ...940..136P}. Compact groups of dwarfs have been studied in the simulations of Small MultiDark Planck by \citet{2023MNRAS.525..415Y} and TNG50 in \citet{2024arXiv240113252F}.
Furthermore, TiNy Titans \citep{2015ApJ...805....2S}, DELVE \citep{2021ApJS..256....2D}, MADCASH \citep{2016ApJ...828L...5C}, and LBT-SONG \citep{2021MNRAS.500.3854D,2024arXiv240903999D,2021MNRAS.507.4764G} have surveyed such systems in the Local Volume. Investigating these systems will lead to a better understanding of their contribution to the hierarchical assembly of the large-scale structure of the universe, and possibly CDM constraints on the smallest scales \citep{2024ApJ...967...61N}.

We do find from Fig. \ref{fig:env_dssfr} and \ref{fig:cumulativeSFH} that the distributions of the density estimators, as well as the cumulative SFH, are comparable between the secondaries and primaries. While this shows that the effect of quenching due to the primaries is minimal for secondaries with ${\rm log}(\mathcal{M}_{\ast}/M_{\odot}) \geq 7.5$, there is evidence in the literature of quenching mediated by dwarf-dwarf interactions \citep[e.g.,][]{2015ApJ...805....2S,2016MNRAS.459.1827P,2020MNRAS.492.1713G,2021MNRAS.507.4764G,2023arXiv231109280K,2021ApJ...909..211C}. However, doing a thorough analysis of this for the dwarfs in TNG50 simulation is proposed as a future work.

We do find a divergence in the properties of the secondaries in Fig. \ref{fig:subhalo_mass}. We find that these systems are, by and large, star-forming, showing that the tidal stripping of the dark matter has neither affected star-formation nor significantly reduced their gas-fractions. \citet{2022MNRAS.513.2673J} find tidal structures around satellites with $6 <{\rm log}(\mathcal{M}_{\ast}/M_{\odot})<7$ in simulated LMC-mass groups. There have been observations of tidal structures in LMC/SMC analogs as well \citep{2012ApJ...748L..24M,2019ApJ...886..109C}. With a large enough sample of such tidally stripped satellites of LMC analogs, weak lensing can be used to investigate their dark matter distribution better, e.g., with the star-forming dwarf sample of Merian \citep{2023arXiv230519310L} and Dark Energy Survey \citep{2023arXiv231114659T}, and constrain the galaxy-halo connection at these mass scales \citep{2017MNRAS.467.2019R,2021ApJ...923...35M,2024ApJ...961..236C}.

\section{Conclusion} \label{sec:conclusion}

In this work, we focus on dwarf galaxies in the simulated volume of the TNG50 simulation of the IllustrisTNG project with stellar masses $7.5 <{\rm log}(\mathcal{M}_{\ast}/M_{\odot})<9.5$ and are analogous to the LMC/SMC galaxies. We study star-formation in field dwarfs and show their connection to their large-scale environment. We then study the star-formation of these dwarfs.

\begin{itemize}
    \item The dwarfs which have low host halo mass $9 < {\rm log}(\mathcal{M}_{200}/M_{\odot}) < 11.5$, mainly exist in the field of the simulation volume, within overdensities of ${\rm log}(1+\delta_5)\lesssim 2$ (see Fig. \ref{fig:dens_map}). This is in contrast to the satellite dwarfs occupying the dense environments up to ${\rm log}(1+\delta_5) \sim 6$ near group- and cluster-scale hosts of masses ${\rm log}(\mathcal{M}_{200}/M_{\odot}) \geq 11.5$.
    \item Nonetheless, there is still great diversity among the large-scale environment of the field dwarfs. At fixed environmental density, the field population is less quenched compared to the satellite dwarfs. The phenomenon of starbursts is absent in the field dwarfs and only take place in the satellites dwarfs within massive hosts. 
    \item The extremes of star-formation, i.e., sSFR$> \pm 1$ dex of the star-forming main sequence, is mainly restricted to environments with $\dhst<1.5$ Mpc \citep{2012ApJ...757...85G} and $\tht>0$, containing 88\% and 92\% of the quenched field dwarfs, respectively. These are mostly backsplash galaxies, with a fraction of primary dwarfs that have been processed in the denser environments near the massive galaxies. 
    \item In particular, those dwarfs at $\dhst>1.5$ Mpc and $\tht<0$ are well isolated, and there is little to no environmental trend in their sSFR that coincides with the star-forming main sequence at $\sim 0.1\ {\rm Gyr}^{-1}$. This population of field dwarfs is composed of primaries and secondaries, only $\sim 1\%$ of which are quenched. 
    \item The quenched population is largely made up of backsplash dwarfs that have previously interacted with cluster-scale host halos of ${\rm log}(\mathcal{M}_{200}/M_{\odot}) \gtrsim 13$, as we show in Fig. \ref{fig:quench_mneigh}. There is also a small number of quenched primaries that is in proximity to less massive host halos with $11.5< {\rm log}(\mathcal{M}_{200}/M_{\odot}) \lesssim 13$.
    \item This quenched population is also responsible for a two-halo conformity signal at the scales of $\dhst<1.5$ Mpc (see Fig. \ref{fig:quench_conform}), which is primarily driven by the backsplash dwarfs.
    \item The backsplash and secondary dwarfs are deficient in dark matter, with lower $\mathcal{M}_{\rm dyn}$ at fixed $\mathcal{M}_{\ast}$ and $\mathcal{M}_{\rm gas}$ compared to the isolated primary dwarfs in Fig. \ref{fig:subhalo_mass}. This is a signature of tidal stripping due to the neighboring massive galaxies and dwarf primaries, respectively. While the backsplash dwarfs are found to be gas-poor and quenched, the secondaries are gas-rich and still forming stars.
    
\end{itemize}

To summarize, this study connects the star-formation in the field dwarfs in TNG50 with the local and large-scale environments that they belong to. This is expedient as we realize the importance of the LMC/SMC galaxies in the evolution of the MW, and with the growing focus on massive dwarf galaxies in observational work. This has led to an emphasis to observe such systems and their substructure in the Local Volume, e.g. with DELVE, MADCASH, and LBT-SONG. Weak lensing studies of these dwarfs will enable mass measurement of their dark matter halos \citep{2023arXiv231114659T,2023arXiv230519310L}, which will lead to a better description of their local and large-scale environments. The measurements will further test the connection between star-formation and environment as highlighted in this work. Furthermore, we will benefit from a higher resolution simulation in a larger cosmological volume. To get a wider view on environmental quenching of dwarfs, the simulation should capture star formation reliably in smaller dwarfs at ${\rm log}(\mathcal{M}_{\ast}/M_{\odot})<7.5$, as well as containing a sufficient number of massive clusters at ${\rm log}(\mathcal{M}_{200}/M_{\odot})>14$. This will also enable us to model the abundances, orbits and star-formation histories of a larger number of backsplash dwarfs. Finally, the theoretical expectations can be tested with the datasets arising from the surveys of the current generation, including DESI \citep{2023ApJ...954..149D} and LSST \citep{2009arXiv0912.0201L}. These will provide a better understanding of the connection among dwarf galaxies, their star-formation, and environment.

\section{Acknowledgment}

The authors thank Alyson Brooks and the anonymous referee for comments that were highly valuable for the study. JB would like to thank Kavli Institute of Theoretical Physics (KITP), for their hospitality afforded during her stay. This research was supported in part by grant NSF PHY-2309135 to the Kavli Institute for Theoretical Physics (KITP). \\

Software: \texttt{NumPy} \citep{2020Natur.585..357H}, \texttt{Matplotlib} \citep{2007CSE.....9...90H},\texttt{SciPy} \citep{2020SciPy-NMeth}, \texttt{Astropy} \citep{2013A&A...558A..33A,2018AJ....156..123A}, \texttt{COLOSSUS} \citep{2018ApJS..239...35D}.

\bibliography{sample631}{}
\bibliographystyle{aasjournal}

\appendix
\section{Effect of Massive Host Halos} \label{sec:massive_host}

\begin{figure*}
\centering
\includegraphics[scale=0.35]{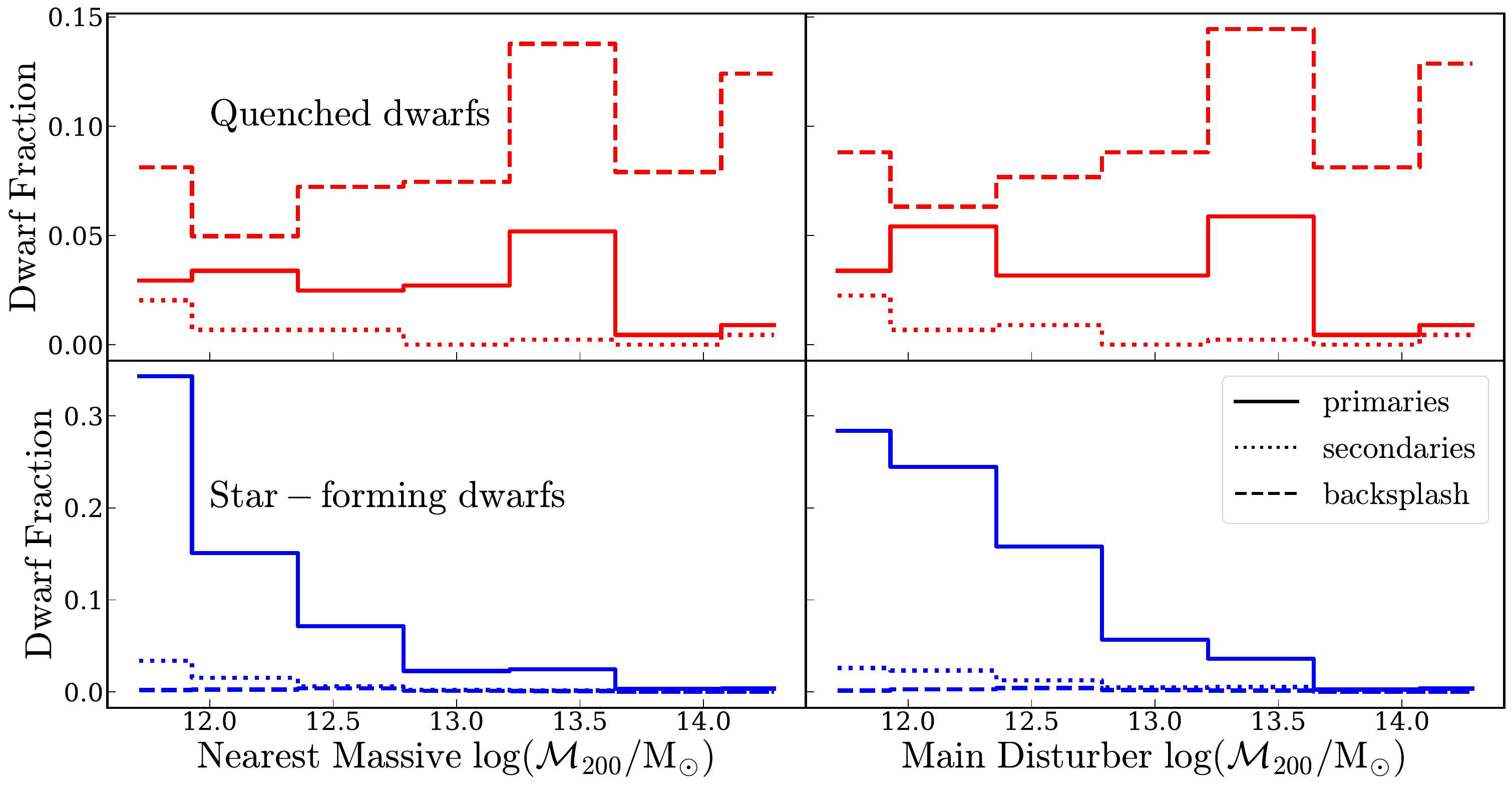}
\caption{Histograms of the host halo mass ${\rm log}(\mathcal{M}_{200}/M_{\odot})$ of the field dwarfs' neighboring Nearest Massive galaxy \textit{(left column)} and the Main Disturber \textit{(right column)}. The \textit{upper row} shows only the quenched field dwarfs and the \textit{lower row} shows the star-forming field dwarfs. These are normalized by the total quenched and star-forming dwarfs, respectively. The primaries, backsplash dwarfs, and secondaries are represented using \textit{solid, dashed} and \textit{dotted} lines, respectively. Among the quenched dwarfs, the backsplash galaxies are prevalent as neighbors of cluster-scale hosts with ${\rm log}(\mathcal{M}_{200}/M_{\odot})\gtrsim 13$. On the other hand, the star-forming field dwarfs are found to be associated with the group-scale hosts with $11.5< {\rm log}(\mathcal{M}_{200}/M_{\odot})< 13$. \label{fig:quench_mneigh}}
\end{figure*}

Having identified the quenched population of backsplash and primary dwarfs in the denser environment, we investigate the masses of their neighboring massive galaxies here. In Fig. \ref{fig:quench_mneigh}, we plot the histograms for host halo masses ${\rm log}(\mathcal{M}_{200}/M_{\odot})$ of the nearest massive galaxy and the Main Disturber that determined the calculation of $\dhst$ and $\tht$, respectively. This is done for two cases---ones of the quenched field dwarfs and the star-forming field dwarfs, respectively, that are shown in the upper and lower panels of the figure, respectively. These histograms in the two instances are normalized by the total number of quenched and star-forming dwarfs in the field, respectively. We show the histograms for the classes of primaries, backsplash, and secondaries shown using the solid, dashed, and dotted lines, respectively.

Among the quenched field dwarfs, we had earlier seen that these were mainly backsplash with a few primaries that are located at $\dhst<1.5$ Mpc and $\tht>0$. From the histograms in the upper panel of Fig. \ref{fig:quench_mneigh}, we read that the most of the backsplash dwarfs are found near the cluster-scale massive hosts with ${\rm log}(\mathcal{M}_{200}/M_{\odot})\gtrsim 13$. However, the corresponding distributions for the primaries and secondaries do not show any notable trend except for them being suppressed in two of the highest mass bins. In the lower panel, we find that among the star-forming dwarfs, the primaries which are most numerous here have increasing abundance with decreasing neighboring host mass across $11.5< {\rm log}(\mathcal{M}_{200}/M_{\odot})< 13$. Meanwhile, the star-forming secondary and backsplash dwarfs are significantly smaller in population compared to the primaries.

\section{A broader definition of the `field'} \label{sec:outside_halo}

Throughout this work, we have used a sample of field dwarf galaxies that reside in SUBFIND-assigned FoF groups with halo masses $9 < {\rm log}(\mathcal{M}_{200}/M_{\odot}) < 11.5$. This is demonstrated in Fig. \ref{fig:dens_map}, wherein we find that with this sample, we are selecting a clean sample of dwarfs in the low-density environment of TNG50. However, there also exists an alternative definition of the field, that considers galaxies which are beyond a certain distance from a massive galaxy to be in the field. The virial radius $R_{200}$ is commonly used as a metric of the extent of a massive galaxy, and thereby this is used to demarcate the field galaxies from the satellite galaxies around it in both simulations and observations alike \citep[e.g.][]{2014MNRAS.438.2578G,2017ApJ...847....4G,2022ApJ...933...47C,2023MNRAS.522.5946E}.

\begin{figure*}
\centering
\includegraphics[scale=0.375]{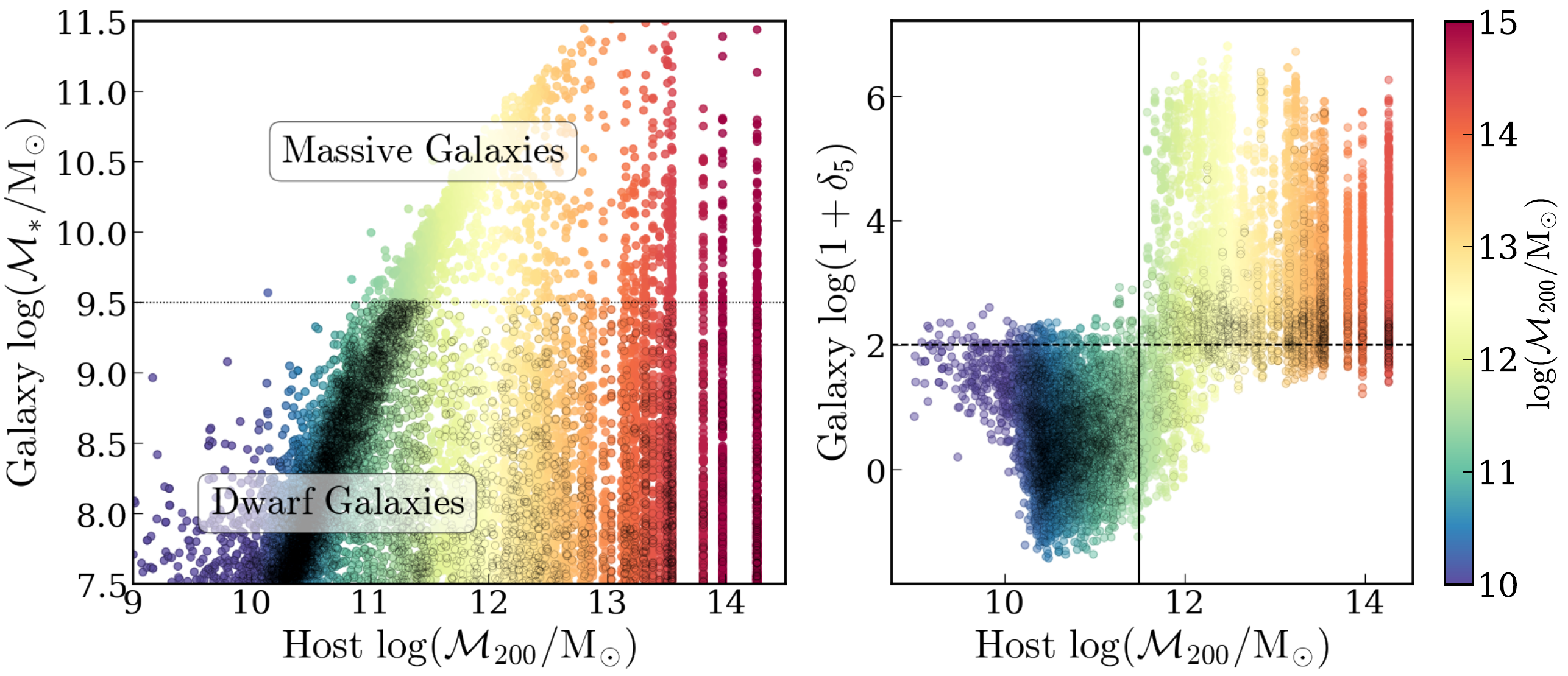}
\caption{\textit{Left:} An iteration of the Fig. \ref{fig:hostM_subhaloM} scatter plot showing the stellar mass $\mathcal{M}_{\ast}$ vs. the host virial mass $\mathcal{M}_{200}$ for all the galaxies in the box of TNG50. We depict the extended sample of field dwarf galaxy using the points with edges. \textit{Right:} An iteration of Fig. \ref{fig:dens_map}, showing the galaxies in  $\mathcal{M}_{200}-{\rm log}(1+\delta_5)$ space, where the vertical solid line defines the limit ${\rm log}(\mathcal{M}_{200}/M_{\odot}) = 11.5$ used to define the fiducial field dwarf sample on its left, whereas the additional dwarfs in the extended sample are its right. 
\label{fig:Bdens_map}}
\end{figure*}

Here, we present versions of the plots in the main text for an extended sample of field dwarf galaxies. This sample consists of the fiducial sample,  supplemented with dwarf galaxies that reside in SUBFIND-assigned hosts with masses ${\rm log}(\mathcal{M}_{200}/M_{\odot}) > 11.5$ yet are beyond their virial radius at distances $d>R_{200}$ from the centers of their hosts. For a backsplash dwarf, $d/R_{200}$ corresponds to the $z=0$ distance to the center of the massive halo inside which it completed its pericentric passage, normalized with respect to the $R_{200}$ of the same halo. With this prescription, we add 1437 dwarfs to the fiducial sample of size 5843. This addition includes 882 backsplash and 555 primary dwarfs that represent almost two-fold and 11\% increase in the size of the sub-samples respectively. However, we were unable to discern if these primaries were satellites of a massive galaxies but SUBFIND does not provide this information. In Fig. \ref{fig:Bdens_map}, we show the distributions of this sample in versions of Fig. \ref{fig:hostM_subhaloM} and \ref{fig:dens_map}. The added population of dwarfs are situated beyond the virial volume of the massive halos with ${\rm log}(\mathcal{M}_{200}/M_{\odot}) > 11.5$, and occupy the denser large-scale environments of the box with ${\rm log}(1+\delta_5) \approx 2$.

\begin{figure*}
\centering
\includegraphics[scale=0.35]{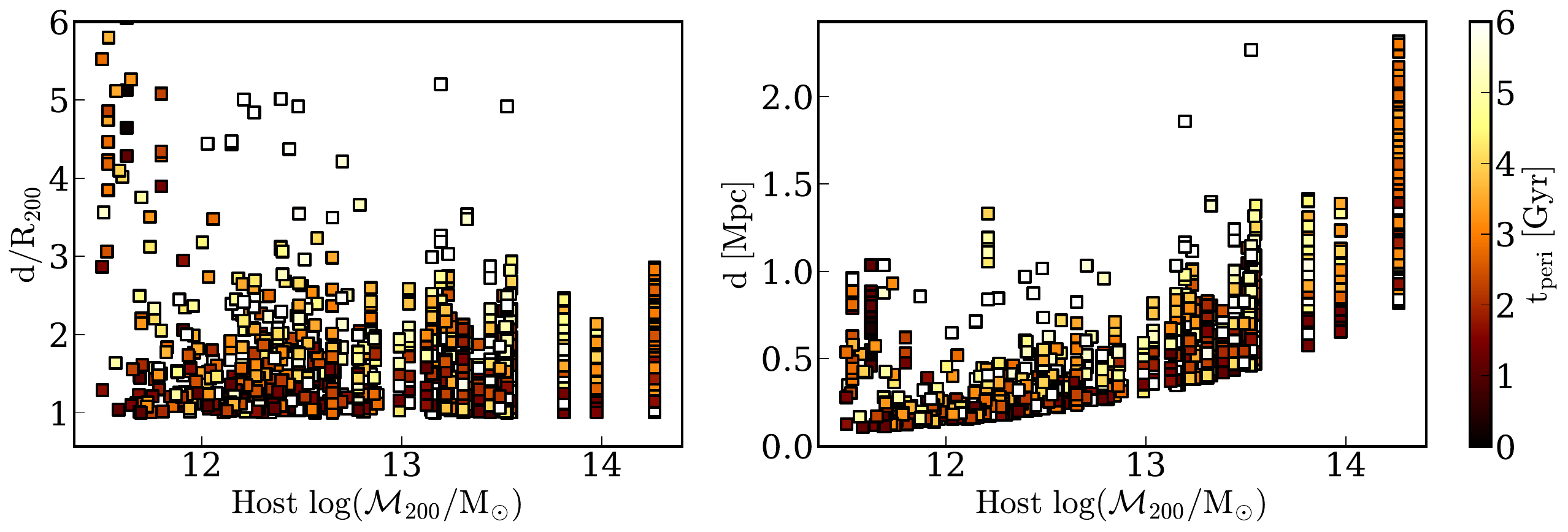}
\caption{ A version of Fig. \ref{fig:backspl_rads} that shows the radial distribution of backsplash dwarfs in the extended field sample, with present day distances normalized with respect to the virial radius of the massive halo where it had its last pericentric passage (\textit{left}) and present day distances in units of Mpc (\textit{right}). A total of 882 backsplash dwarfs have been added to our sample that reside outside the virial radius of massive hosts with masses ${\rm log}(\mathcal{M}_{200}/M_{\odot}) \sim 12-13$. \label{fig:Bbackspl_rads}}
\end{figure*}

We find it noteworthy that upon extending our field sample to include the denser regions in proximity to the massive halos, we add a sample of dwarfs of which about 60\% are backsplash galaxies. The radial distributions of these galaxies are depicted in Fig. \ref{fig:Bbackspl_rads}. The additional systems are dominantly at low $d/R_{200}$, and we add relatively more for low host masses than for higher host masses. In all, we notice that among all the backsplash dwarfs in the simulation, we select a fraction, i.e., 30\% of them in the fiducial sample. Since we determine the fiducial sample using their halo mass, only those backsplash galaxies which are in relatively under-dense regions are assigned their own FoF groups by SUBFIND, enter into our sample. In contrast, those backsplash systems that are in the relatively dense regions around the massive hosts with halo masses of ${\rm log}(\mathcal{M}_{200}/M_{\odot}) \sim 12-13$ are assigned as the latter's subhalos. We leave a detailed investigation of these systems, that account for the full sample of the backsplash galaxies across all halos in different large-scale environments in TNG50, for a future work.

\begin{figure*}
\centering
\includegraphics[scale=0.275]{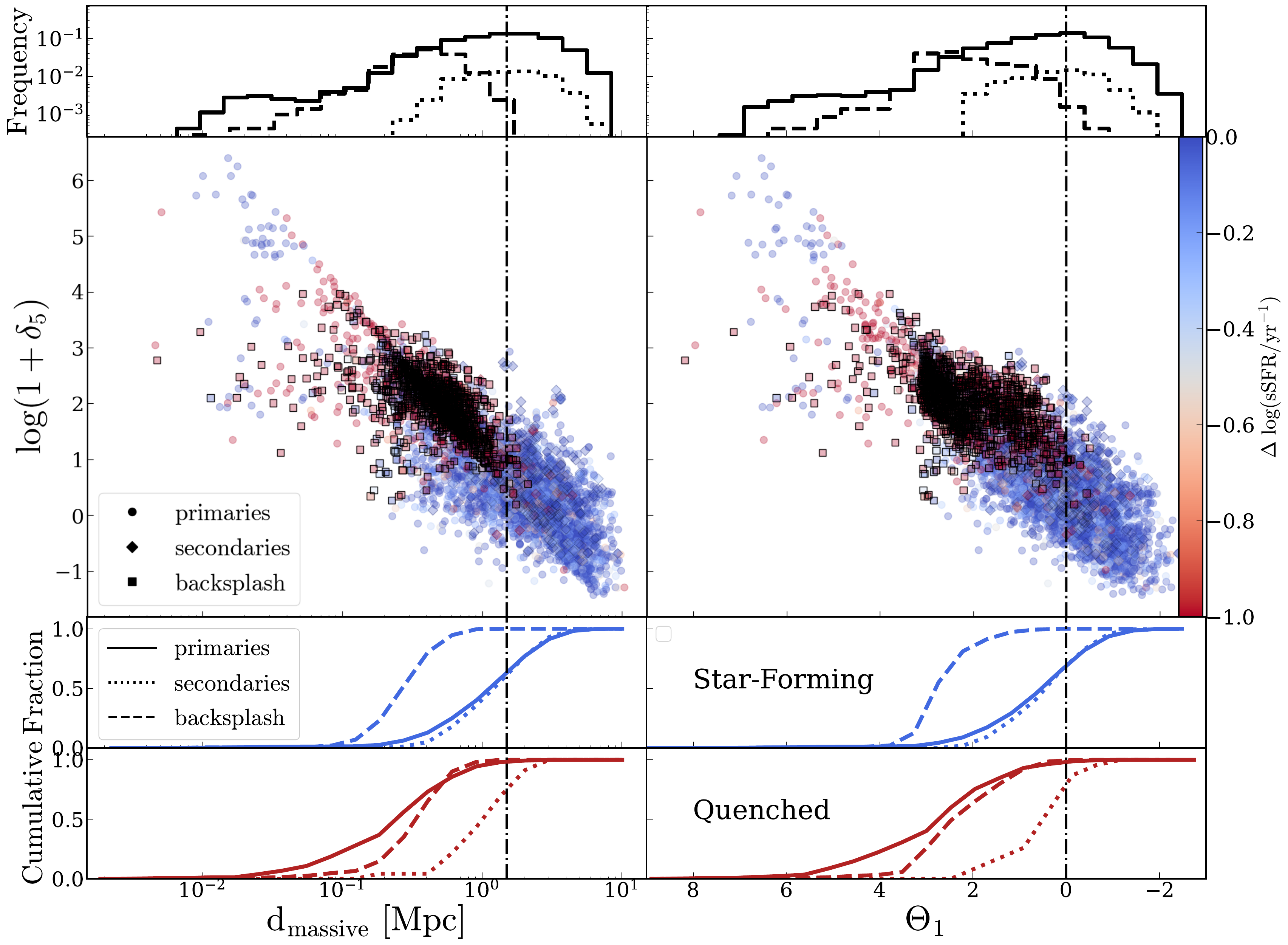}
\caption{The distributions of the field dwarf sample in $\dhst-{\rm log}(1+\delta_5)$ (\textit{left}) and $\tht -{\rm log}(1+\delta_5)$ (\textit{right}) spaces, and is to be compared to Fig. \ref{fig:env_dssfr}. In the \textit{lower} panel we plot the cumulative fractions of star-forming and quenched field dwarfs, respectively, with \textit{blue} and \textit{red} lines, respectively. Upon including the additional dwarfs, the scatter in ${\rm log}(1+\delta_5)$ increases in the regions of $\dhst<1.5$ Mpc and $\tht>0$. \label{fig:Benv_dssfr}}
\end{figure*}

\begin{figure*}
\centering
\includegraphics[scale=0.3]{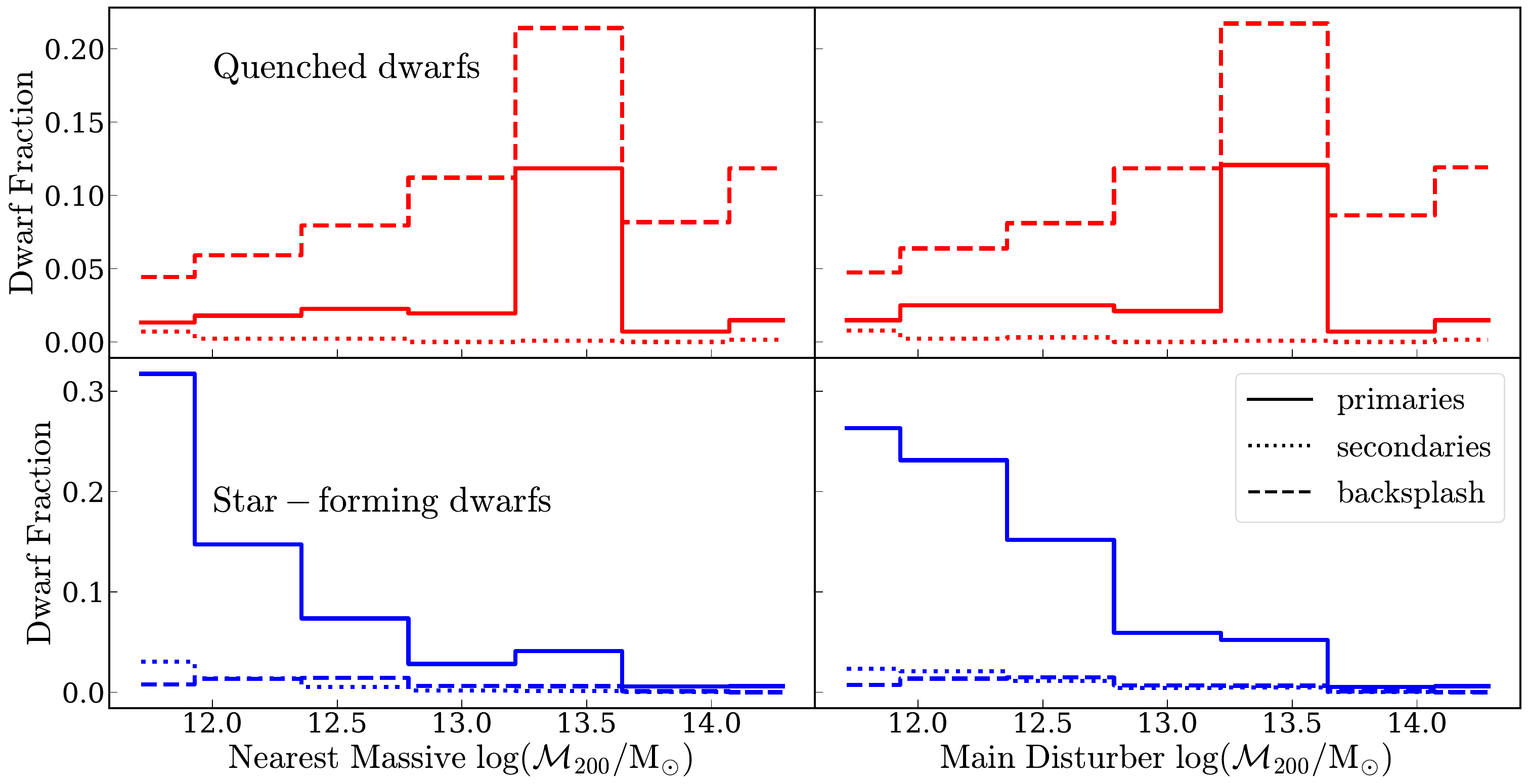}
\caption{Histograms of the host halo mass ${\rm log}(\mathcal{M}_{200}/M_{\odot})$ of the extended field dwarfs' neighboring- Nearest Massive galaxy \textit{(left column)} and the Main Disturber \textit{(right column)} and is to be compared to Fig. \ref{fig:quench_mneigh}. The \textit{upper row} shows only the quenched field dwarfs and the \textit{lower row} shows the star-forming field dwarfs. These are normalized by the total quenched and star-forming dwarfs respectively. The quenched backsplash and primary dwarfs are prevalent as neighbors of cluster-scale hosts with ${\rm log}(\mathcal{M}_{200}/M_{\odot})\gtrsim 13$. \label{fig:Bquench_mneigh}}
\end{figure*}

In Fig. \ref{fig:Benv_dssfr}, we find that the extended field sample has a large scatter in the high-density regions of the parameter space, mainly in terms of the overdensity parameter, which is now spread over $-1.43<{\rm log}(1+\delta_5)<6.39$, with a median of 0.8. With the inclusion of the backsplash and infalling dwarfs with $d/R_{200} >1$ in the massive halo, the overall quenched fraction also increases to 17\% as opposed to 8\% for the fiducial sample. However, 97\% of quenched dwarfs are situated at $\dhst<1.5$ Mpc and $\tht>0$.  Of the dwarfs with $\dhst>1.5$ Mpc and $\tht<0$, only 1\% are quenched. This is in close agreement with the relevant estimates from the fiducial sample, and shows that our results are robust to the definition of the field dwarf sample. In Fig. \ref{fig:Bquench_mneigh}, we show the host mass distributions for the neighbors of the extended field sample, which is comparable to Fig. \ref{fig:quench_mneigh}, especially in the fact that the quenched backsplash and primary dwarfs tend to be found around the cluster-mass hosts with ${\rm log}(\mathcal{M}_{200}/M_{\odot})\gtrsim 13$.

\begin{figure*}
\centering
\includegraphics[scale=0.3]{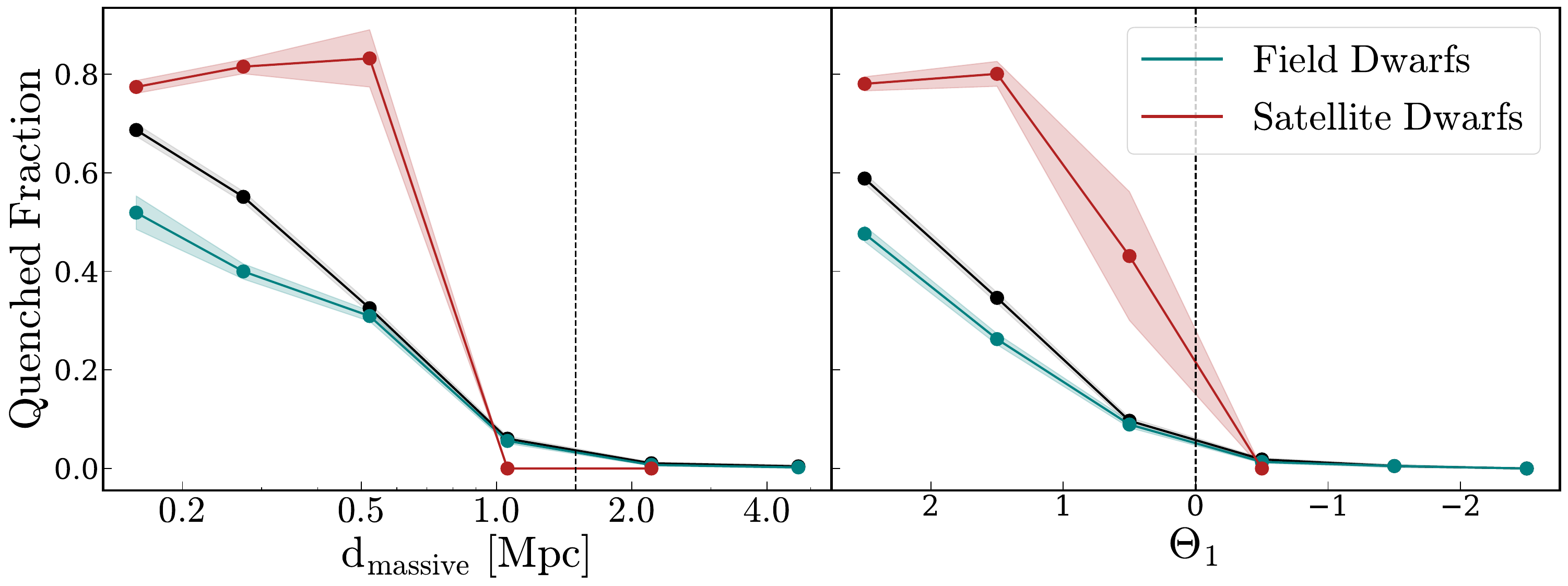}
\includegraphics[scale=0.3]{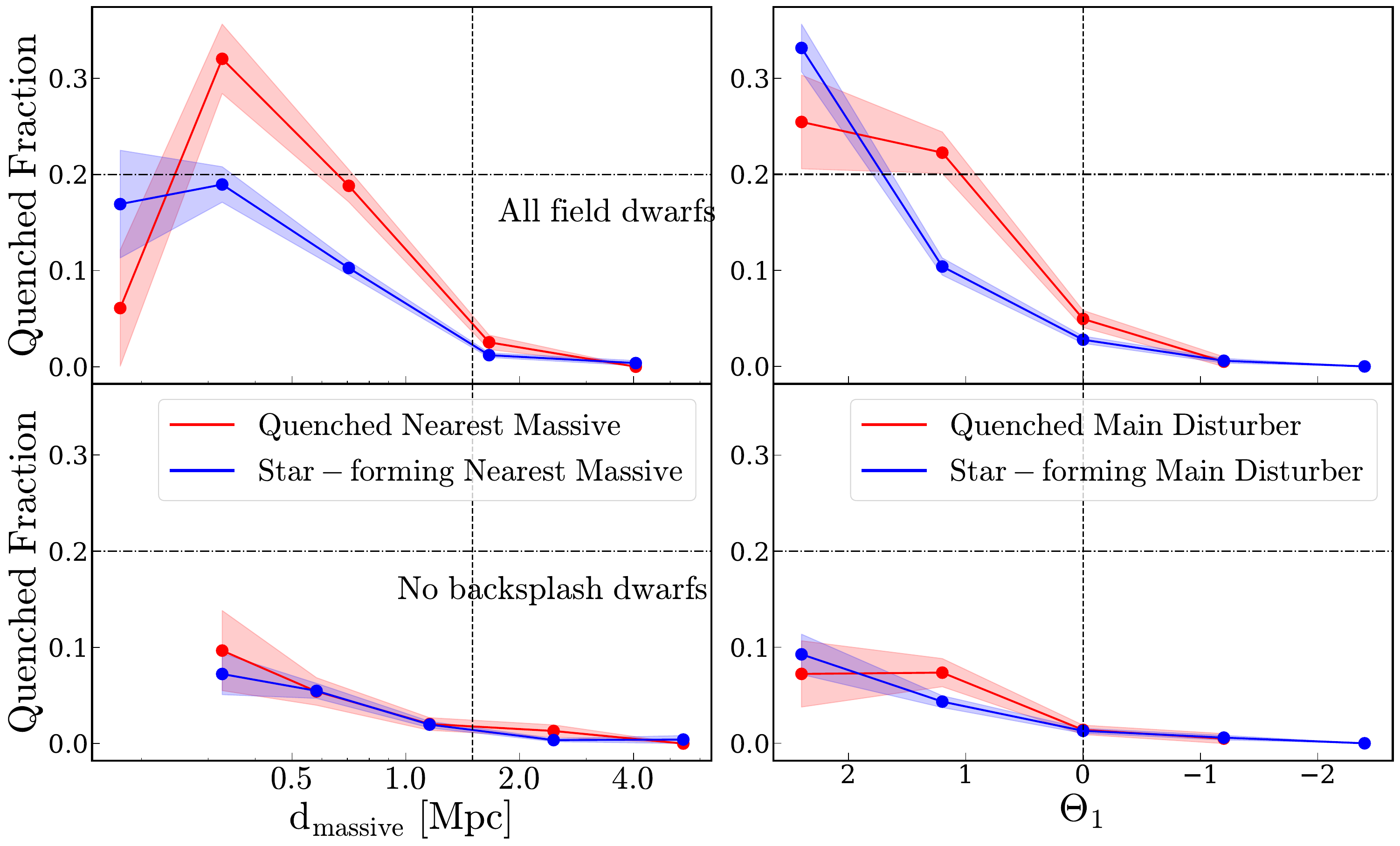}
\caption{\textit{Upper row:} The quenched fraction for all the dwarf galaxies in the TNG50 box, the extended field and the satellite samples in bins of $\dhst$ and $\tht$ that is to be compared to Fig. \ref{fig:quench_tracer}. \textit{Center row:} The quenched fraction for the field dwarf sample in bins of $\dhst$ and $\tht$- \textit{left:} the distance to the nearest massive galaxy $\dhst$,  \textit{right:} the tidal index $\tht$, and is to be compared with \ref{fig:Bquench_conform}. The red and blue profiles respectively represent the field dwarfs selected around quenched and star-forming massive galaxies. While in the \textit{upper left} panel we find the effect of two-halo galactic conformity, when we remove the backsplash dwarfs from our sample the difference in the quenched fractions is removed in the \textit{lower left} panel. \label{fig:Bquench_conform} }
\end{figure*}

The uppermost panel of Fig. \ref{fig:Bquench_conform} depicts the dependence of the quenched fractions on $\dhst$ and $\tht$ further. We find that in the densest regions, the quenched fraction increases to about 50\% with the extension of the field dwarf sample. Nonetheless, the influence of the local and large-scale environment on the quenched fraction that we inferred from the fiducial sample in Fig. \ref{fig:quench_tracer} remains the same. Furthermore, in the central and lower panels of Fig. \ref{fig:Bquench_conform}, we show how the two-halo conformity effect that we observed in Fig. \ref{fig:quench_conform} manifests in the extended field sample. We find that the quenched fractions are higher near the quenched massive galaxies and vice-versa in the full sample. When we remove the backsplash dwarfs from the sample, this trend is suppressed, showing that not only is the two-halo conformity present, but also that it is driven largely by the backsplash systems.

\end{document}